\documentclass[11pt,twoside,a4paper]{article}
\usepackage{slashbox}
\usepackage{braket}

\newcommand{\mathsym}[1]{{}}

\newcommand{\baz}{\begin{array}{cc}}
\newcommand{\bad}{\begin{array}{ccc}}
\newcommand{\bi}{\begin{itemize}}
\newcommand{\ei}{\end{itemize}}
\newcommand{\ba}{\begin{array}{c}}
\newcommand{\ea}{\end{array}}

\newcommand{\beqa}{\begin{eqnarray}} 
\newcommand{\eeqa}{\end{eqnarray}} 

\usepackage[utf8]{inputenc}
\usepackage[hang,normal,footnotesize]{caption}
\usepackage{subcaption}
\usepackage{wrapfig}
\usepackage[sort&compress,numbers,colon,merge]{natbib}

\usepackage{a4wide}
\usepackage{fancyhdr}
\usepackage{graphicx}
\usepackage{url}
\usepackage{amsfonts}
\usepackage{amsmath,amssymb,amsthm}

\usepackage{mathrsfs}

\usepackage{multirow}
\usepackage{color}
\definecolor{orange}{rgb}{1,0.5,0}
\definecolor{darkred}{rgb}{0.5,0,0}
\definecolor{darkgreen}{rgb}{0,0.5,0}
\definecolor{darkblue}{rgb}{0,0,0.5}

\usepackage{hyperref}
\hypersetup{ colorlinks,
linkcolor=darkgreen,
urlcolor=darkred,
citecolor=darkblue }

\usepackage{cancel}
\usepackage{bbm}
\usepackage{ulem}
\usepackage{color}

\usepackage{units}

\def\be{\begin{equation}}
\def\ee{\end{equation}}

\newcommand{\bea}{\begin{equation} \begin{array}{c}}
\newcommand{\eea}{ \end{array} \end{equation}}

\def\gs{\mathrel{
   \rlap{\raise 0.511ex \hbox{$>$}}{\lower 0.511ex \hbox{$\sim$}}}}
\def\ls{\mathrel{
   \rlap{\raise 0.511ex \hbox{$<$}}{\lower 0.511ex \hbox{$\sim$}}}}

\numberwithin{equation}{section}

\begin{document}

\begin{titlepage}

% Preprint numbers

\begin{center}
 {\huge\sffamily\bfseries\mathversion{bold} 
Naturalness, Vacuum Stability and Leptogenesis in the Minimal Seesaw Model}
\\[5mm]
{\large
Gulab Bambhaniya\footnote{\texttt{gulab.bambhaniya@gmail.com }}$^{(a)}$,~P.\ S.\ Bhupal Dev\footnote{\texttt{bdev@wustl.edu}}$^{(b, c)}$,~Srubabati Goswami\footnote{\texttt{sruba@prl.res.in}}$^{(a)}$,~\\
Subrata Khan\footnote{\texttt{khansubrata@gmail.com }}$^{(d,e)}$,~Werner Rodejohann\footnote{\texttt{werner.rodejohann@mpi-hd.mpg.de}}$^{(b)}$\\
}\mbox{ } \\%[5mm]
%\\[5mm]
{\small\textit{$^{(a)}$Physical Research Laboratory, Navrangpura, Ahmedabad 380009, India
}}\\%[.5cm]
{\small\textit{$^{(b)}$Max-Planck-Institut f\"ur Kernphysik, Saupfercheckweg 1, 69117
Heidelberg, Germany
}}\\%[.5cm]
{\small\textit{$^{(c)}$Department of Physics and McDonnell Center for the Space Sciences, 
\newline
Washington University, St. Louis, MO 63130, USA
}}\\%[.5cm]
{\small\textit{$^{(d)}$Department of Physics, University of Calcutta, Kolkata 700009, India 
}}\\
{\small\textit{$^{(e)}$Department of Physical Sciences, Indian Institute of Science Education and Research  Kolkata, Mohanpur, Nadia 741246, India
}}\\[.5cm]

\end{center}
\vspace*{1.0cm}

\begin{abstract}
\noindent
The right-handed neutrinos within the type-I seesaw mechanism 
can induce large radiative corrections to the Higgs mass, and 
naturalness arguments can then be used to 
set limits on their mass scale and Yukawa couplings.  
Driven by minimality, we consider the presence of two 
degenerate right-handed neutrinos. We compare the limits from 
naturalness with the ones from the stability of the electroweak vacuum 
and from lepton flavor violation. Implications from 
neutrinoless double beta decay are also discussed and 
renormalization effects for the light neutrino parameters are presented. 
Adding small perturbations to the degenerate heavy neutrino spectrum allows for
successful leptogenesis. 
\end{abstract}

\end{titlepage}

\setcounter{footnote}{0}

%%%%%%%%%%%%%%%%%%%%%%%%%%%%%%%%%%%%%%%%%%%%%%%%%%%%%%%%%%%%%%%%%%%
%\keywords{Beyond Standard Model, Vacuum stability, Lepton Flavor Violation, Neutrino Physics}
%%%%%%%%%%%%%%%%%%%%%%%%%%%%%%%%%%%%%%%%%%%%%%%%%%%%%%%%%%%%%%%%%%%
%\preprint{PRL-XXXXX}
\section{Introduction}
The journey of experimentally verifying the 
particle content of the Standard Model (SM) of particle physics has
finally been  completed  by the discovery of the Higgs boson in 2012~\cite{Aad:2012tfa,Chatrchyan:2012xdj}. This seems to clarify the origin of masses for all  Standard Model (SM) particles. 
%All laboratory searches for physics beyond the SM so far have only confirmed the SM to ever-increasing precision. 
On the other hand, the discovery of neutrino oscillations~\cite{Kajita:2016cak} and flavor conversion~\cite{McDonald:2016ixn} require at least two neutrinos to be massive, which cannot be explained in the SM and necessitate the existence of new particles. It is therefore a potentially rewarding problem to investigate the connection between neutrino mass and the Higgs sector of the SM. 
There are indeed several possible non-trivial consequences of 
neutrino mass physics on the SM Higgs boson, such as 
\begin{itemize}
\item  {\it Non-standard Higgs decays:} These could be either into new particles that are implied by models for neutrino mass or lepton mixing, or modified/new decays into SM particles; see  Refs.~\cite{Bhattacharyya:2010hp, Dev:2012zg, Cely:2012bz, Bandyopadhyay:2012px, Banerjee:2013fga, Arganda:2014dta, Hessler:2014ssa, Campos:2014zaa,  Heeck:2014qea,  Antusch:2015mia, Dermisek:2015vra,  Maiezza:2015lza, Kobayashi:2015gwa, Bonilla:2015jdf} 
for some examples. 

\item {\it Vacuum stability:} The couplings of the Higgs boson to new particles modify
  the renormalization group (RG) running of the Higgs self-coupling, and hence, the stability of the electroweak vacuum as compared to that in the SM~\cite{Buttazzo:2013uya}; see for instance Refs.~\cite{Casas:1999cd, Gogoladze:2008ak, EliasMiro:2011aa, He:2012ub, Rodejohann:2012px, Chakrabortty:2012np, Masina:2012tz, Chao:2012mx, Khan:2012zw,  
  Dev:2013ff, Ho:2013hia, Kobakhidze:2013pya, Datta:2013mta, Chakrabortty:2013zja,  Mohapatra:2014qva, Hamada:2014xka, Bambhaniya:2014kga, Bambhaniya:2014hla, Salvio:2015cja, Rose:2015fua, Lindner:2015qva, Haba:2016zbu}. 

\item {\it Naturalness:} Loop corrections to the Higgs mass are typically quadratic in the mass of the heaviest particle in the loop, a property responsible for the hierarchy problem. Thus, if some heavy particle responsible for neutrino mass couples to the Higgs boson, it could also lead to unacceptably large contributions to the Higgs mass. This sets an {\it upper} limit on the mass scale of the new particles associated to neutrino mass; see e.g.\ Refs.~\cite{Vissani:1997ys, Casas:2004gh, Abada:2007ux, Farina:2013mla, Clarke:2015gwa,
Fabbrichesi:2015zna, Clarke:2015hta, Chabab:2015nel, Clarke:2016jzm, Salvio:2016vxi}. 
\end{itemize}

In this paper, we focus on the latter two aspects, and study the impact of
naturalness and vacuum stability within a type-I seesaw model~\cite{Minkowski:1977sc, Mohapatra:1979ia, Yanagida:1979as, GellMann:1980vs, Glashow:1979nm}, 
where heavy right-handed neutrinos are the new particles responsible for light neutrino masses.  
We will study here the most minimal seesaw realization with only two heavy neutrinos~\cite{Smirnov:1993af, Ma:1998zg, King:2002nf, Kuchimanchi:2002fi, Frampton:2002qc, Raidal:2002xf,  Raby:2003ay,  Dutta:2003ps, Ibarra:2003up, Ibarra:2005qi}; for a review, see e.g.~\cite{Guo:2006qa}. Thus, one light neutrino is massless at the tree level.
Further assuming the masses of the 
heavy neutrinos to be (quasi)degenerate leaves us with 3 free parameters
beyond the directly measurable light neutrino parameters, namely the
mass scale of the heavy right-handed neutrinos plus one complex angle in the Casas-Ibarra (CI) parametrization~\cite{Casas:2001sr}. We use 
naturalness arguments to set limits on these parameters, which 
mostly constrain the new mass scale and the imaginary part of the 
complex angle. These limits are compared with constraints from the stability of the electroweak vacuum and from lepton flavor violation (LFV). We also study the phenomenological implications of this scenario for neutrinoless double beta decay ($0\nu\beta\beta$) and the RG evolution of the light neutrino parameters. 
In addition, we show that the observed baryon asymmetry of the 
Universe (BAU) can be successfully explained, leptogenesis, 
for the parameter values allowed by all other constraints, if small perturbations are added to the heavy mass spectrum.  
%RG effects break the mass degeneracy between the 
%two right-handed neutrinos and allow for successful 
%leptogenesis \cite{GonzalezFelipe:2003fi,Turzynski:2004xy,Branco:2005ye}. We show that within 
%the parameter space that is allowed by naturalness indeed the correct
%baryon asymmetry can be generated. 
%{\bf WR: leptogenesis comment here. }
%Add refs on minimal seesaw (Ibarra?Xing?)?}

The rest of the paper is organized as follows: 
In Section~\ref{sec:mod}, we review the minimal seesaw model and set up our notation. In Section~\ref{sec:nat}, we discuss the constraints obtained on the model parameters from the  naturalness criterion. 
In Section~\ref{sec:cons}, we compare the naturalness constraints with those from electroweak vacuum (meta)stability and LFV. RG effects on the light
neutrino parameters are studied in Section~\ref{sec:rg}. In Section~\ref{sec:yb}, we focus on leptogenesis aspects and conclude in Section~\ref{sec:concl}.  
%Lepton Flavour Violation (LFV) and also discuss the combined 
%constraints on the parameters from naturalless, metastabilty and LFV. 
%Finally we also discuss the implication for neutrino less double beta decay
%and end with the conclusions.  

\section{The Minimal Seesaw Model 
\label{sec:mod}}
%{\bf WR: please check carefully notation and convention with respect
 % to rest of the paper; in particular, write matrices here in bold
  %face, as in rest of paper. However, as the journal will remove this
  %anyway, better to remove bold-faced matrices from the draft; 
%comment on effect or our choice of $\zeta = \pm 1$}.
%\ggb{$\zeta = \pm 1$ is not making any change in $\delta\mu^2$.}

%\subsection{Conventions and Formalism}
We consider the minimal type-I seesaw model with two heavy right-handed 
Majorana neutrinos $N_{R_j}$ (with $j=1,2$),  
%having lepton number $+1$, 
defined by the Lagrangian
\begin{eqnarray}
-{\cal{L}} \  = \ Y_{\nu_{j l}}  \overline{N}_{R_j} \tilde{\phi}^\dag L_{l} 
 + \frac{1}{2} \overline{N}_{R_j} {M_{jk}} N_{R_k}^c     
+       {\rm H.c.}  , 
\label{lag:minimal_type1_seesaw}
\end{eqnarray}
where $  L_l = (\nu_l\, , l)_L^T  $ (with $  l = {e, \mu, \tau}$) is the $SU(2)_L$ lepton doublet and $\tilde{\phi}=i\sigma_2\phi^*$ (with $\sigma_2$ being the second Pauli matrix and $\phi=(\phi^+\, , \phi^0)^T $ being the SM Higgs doublet). The sum over repeated indices is implicitly assumed throughout, unless otherwise specified.  
We consider the scenario in which the two heavy neutrinos are degenerate 
so that the heavy-neutrino Majorana mass matrix is trivial:  
%our case we will assume that only two $N_R$ are present, 
$M = {\rm diag}(M_N,M_N)$. 
% Here both $l_L$ and $N_R$ have lepton number +1.
In general, the neutrino mass matrix in the $(\nu_L,N_R^c)$ basis can be
written as 
\be
M_\nu = 
\bad
\begin{pmatrix}
0 & m_D^T  \\
m_D & M \\
\end{pmatrix}  .
\end{array} 
\label{massmatrix} 
\ee
Here $m_D = v \, Y_\nu/\sqrt{2}$, where $v$ is the vacuum expectation value (VEV) of the SM Higgs field which in our convention is defined as $ \braket{\phi^0} =
v/\sqrt{2} = 174 ~\mbox{GeV}$. 
%{\bf WR: please give relation  between $\mu^2$, v and $m_h$.}
% Now defining $M_R$ as
% \begin{equation}
% M_R = 
% \left(\begin{array}{cc} 
% M_N & 0 \\
% 0  & M_N 
% \end{array}\right),
% \label{msinglet} 
% \end{equation}
The mass matrix can be diagonalized
by a $5\times5$  unitary matrix $ U_0 $ as
\begin{eqnarray}
\label{diagonal}
U_0^T  M_\nu \, U_0  \ = \ M_\nu^{\rm diag} \ = \ \mbox{diag}(m_1,m_2,m_3, M_1,M_2)\,,
\end{eqnarray}
with mass eigenvalues $m_i$ ($i=1,2,3$) and $M_j$ ($j=1,2$) for 
light and heavy neutrinos, respectively. In our case, $M_1 = M_2 = 
M_N$,  $m_1=0$ for the normal hierarchy (NH), and $m_3=0$ for the inverted hierarchy (IH). 
We can write $U_0$ as~\cite{Schechter:1981cv, Korner:1992zk, Grimus:2000vj, Hettmansperger:2011bt, Khan:2012kc, Dev:2012sg}
\begin{eqnarray}
U_0 \ = \ W\, U_{\nu} \ \simeq \ 
\left(\begin{array}{cc}  
\left( \mathbbm{1}-\frac{1}{2}\epsilon \right ) U & {m_D^{ \dagger}} (M^{-1})^{\ast} U_{R}\\
-M^{-1}m_D^{} U & \left(\mathbbm{1}-\frac{1}{2}\epsilon'\right) U_{R}
\end{array} \right) 
 \equiv \left(\begin{array}{cc}
U_{L} & T\\
S & U_{H }
 \end{array} \right) ,
\label{bdmatrix} 
\end{eqnarray}
where $W$ is the matrix which block-diagonalizes the 
full  $5 \times 5$ neutrino matrix: 
\begin{eqnarray}
 W^T \begin{pmatrix}
 0 & m_{D}^{ T}\\
 m_{D}^{ } & M 
\end{pmatrix}W
= \begin{pmatrix}
 m_{\text{light}} & 0\\
 0 & m_{\text{heavy}}
\end{pmatrix}. 
\label{blockdiagonal}
\end{eqnarray}
Here $U_\nu = {\rm diag}(U,U_R)$ diagonalizes the mass matrices in
the light and heavy
sector appearing in the upper and lower block of the block diagonal 
matrix, respectively, in Eq.~\eqref{blockdiagonal}. In our case,  $U_R = \mathbbm{1}$ and $m_{\rm
  heavy} = M$.  The matrix $U_L$ in the upper left corner of Eq.\
(\ref{bdmatrix}) is the new PMNS mixing matrix,  
which acquires a non-unitary correction over the original PMNS matrix $U$, which is the matrix that diagonalizes $m_{\rm light}$. 
% The eigenvalues $(M_1,M_2)$ are obtained 
% as $(-M_N,M_N)$ corresponding to degenerate neutrinos 
% with opposite CP parities. 
% The  negative sign in the mass eigenvalues 
% can be absorbed in the phases of the diagonalizing matrix $U_R$ giving,   
% \begin{equation} 
% U_R = \frac{1}{\sqrt{2}}
% \begin{pmatrix} 
% i & 1 \\ -i & 1 
% \end{pmatrix}.
% \label{ur} 
% \end{equation} 
Finally, $\epsilon$ and $\epsilon'$ characterize the non-unitarity and 
are given by 
\begin{eqnarray}
\epsilon \ = \ T T^{\dagger} \ = \ m_D^{ \dagger} \left(M^{-1}\right)^\ast M^{-1} m_{D}^{}\,,\nonumber \\
\epsilon' \ = \ S S^{\dagger} \ = \  M^{-1}m_{D}^{}m_D^{ \dagger}\left(M^{-1}\right)^\ast\,.
\label{epsilons} 
\end{eqnarray}
The light neutrino mass matrix, in the limit $M \gg m_D$, is given as
\begin{eqnarray}
m_{\rm light} \ = \ -m_D^{ T} \, M^{-1} m_D \,.
\label{effective-nu-massmatrix}
\end{eqnarray}
%\blue{no minus sign here?}
% Now inserting the expression for $m_D$, the light neutrino mass matrix    
% is the same as that in Eq. (\ref{eq:mlightlinear}).  
% Note that the complete mass matrix %(\ref{eq:linearseesaw}) 
% for the minimal model has 7 phases out of which 5 can be rotated away by redefinition of the 
% fields. Thus there are 2 independent phases in this matrix. 
% We choose the basis in which $M_R$ is real and attach the phases to the elements of 
% $Y_\nu$ and $Y_s$. Since $M_\nu$ for this case is of rank 4,
% there is one zero eigenvalue. Thus one of the light neutrino states is massless and the two 
% remaining masses are completely determined in terms  of the two mass squared differences measured 
% in oscillation experiments.
% {\color{red}Copied from Subrata Khan???????????????????????????????}
It proves very useful to introduce the Casas-Ibarra parametrization 
for the Yukawa coupling matrix \cite{Casas:2001sr}: 
\begin{equation} \label{ynu}
{Y}_\nu \ = \ \frac{\sqrt{2}}{v} \sqrt{D_{N}} \: R \: \sqrt{D_{\nu}} \: U^\dagger\,,
\end{equation} 
where $D_{N} = {\rm diag}(M_1, M_2)$, 
$D_{\nu}= {\rm diag}(m_1,m_2, m_3)$, and $R$ is an arbitrary $2\times 3$ orthogonal matrix. In the minimal seesaw model, the light masses are completely fixed by the measured solar and atmospheric mass-squared differences:   
\begin{align} 
\begin{array}{cc}
m_1 \ =\ 0, \quad m_2 \ = \ \sqrt{\Delta m^2_{\rm s}}, \quad 
m_3 \ = \ \sqrt{\Delta m^2_{\rm a}}\, , &  {\rm (NH)} \\
m_1 \ = \ \sqrt{\Delta m^2_{\rm a}}, \quad m_2 \ = \ \sqrt{\Delta m^2_{\rm a} +\Delta
  m^2_{\rm s}}, \quad m_3 \ = \ 0\, . &  {\rm (IH)}
\end{array}  
\end{align}   
%for
%the normal hierarchy (NH, $m_1 < m_2 < m_3$) and ${ D_{\sqrt{m_\nu}} }=
%{\rm diag}(\sqrt{m_1}, \sqrt{m_2}, 0)$ for the inverted hierarchy (IH,
%$m_3 < m_1 < m_2$). 
The matrix $U$ in Eq.~\eqref{ynu} diagonalizing the light neutrino mass
matrix $m_{\rm light}$ is parametrized by three mixing angles $\theta_{ij}$ (with $i,j=1,2,3; i\neq j$), one Dirac phase $\delta$ and one Majorana phase $\alpha$:
\begin{align}
U \ = \ \begin{pmatrix} 
c_{12}c_{13} & s_{12} c_{13} & s_{13}e^{-i\delta} \\
-s_{12} c_{23}-c_{12} s_{23} s_{13}e^{i\delta} & 
c_{12}c_{23}-s_{12}s_{23}s_{13}e^{i\delta} & 
s_{23}c_{13} \\
s_{12}s_{23} - c_{12}c_{23}s_{13}e^{i\delta} & 
-c_{12}s_{23}-s_{12}c_{23}s_{13}e^{i\delta} & 
c_{23}c_{13}
\end{pmatrix} {\rm diag}(e^{-i \alpha},e^{+i \alpha},1) \, ,
\end{align}
where $c_{ij}\equiv \cos\theta_{ij}$, $s_{ij}\equiv \sin\theta_{ij}$. 
For numerical purposes, we will use the $3\sigma$ ranges of the mass-squared differences and mixing parameters from the global-fit of Ref.~\cite{Forero:2014bxa} and vary the $CP$ phases $\delta$ and $\alpha$ between $-\pi$ to $+\pi$, unless otherwise specified.   
%our results are not very sensitive to small variations of these
%numbers.  

%\subsection{Determination of R}
From Eq.~(\ref{ynu}) we have
\begin{eqnarray}
 R_{ij} \ = \ \frac{\left({{Y}_\nu}\,U\right)_{ij}}{\sqrt{M_i\,m_j}}\,\frac{v}{\sqrt{2}}\,,
\end{eqnarray}
where $j\ne 1$ for NH and $j\ne 3$ for IH. 
%Note that this scenario can be considered as  a limiting case of three right-handed
%neutrino model in the limit $M_3 \to \infty$, which gives, for NH, $R_{32}\,,R_{33}=0$ and for IH, $R_{31}\,,R_{32}=0$. Orthonormalization 
%condition gives $R_{31}\,, R_{33}=1$ as well as $R_{11}\,,R_{21}=0$ and $R_{13}\,,R_{23}=0$ for NH, IH respectively.  With this for two right 
%handed neutrino case, 
We can parametrize the matrix $R$ as 
\be 
 R \ = \ \left\{\begin{array}{cc}\begin{pmatrix} 0 & \cos{z} & \zeta \sin{z} \\ 0 & -\sin{z} & \zeta \cos{z}
\end{pmatrix} & {\rm (NH)} \\[0.3cm]
\begin{pmatrix} \cos{z} & \zeta \sin{z} & 0 \\ -\sin{z} & \zeta \cos{z} & 0
\end{pmatrix} & {\rm (IH)} \, .
\end{array}\right.
\ee 
Here $z$ is a complex parameter and $\zeta = \pm 1$, which
however has no influence for our results; so we will use $\zeta =
+1$ from now on.

%We use the $3\sigma$ ranges of these parameters from global 
%oscillation analysis \cite{Forero:2014bxa} . 
%\begin{table}%[t!]
%\begin{tabular}{|c||c|c|c|c|c|c|}
%\hline
%\backslashbox{~\hskip -0pt}{\hskip -0pt ~~}
%&$\Delta_{\odot}^2~[10^{-5}\,\textrm{eV}^2]$  &     $\Delta_{\textrm{atm}}^2~[10^{-3}\,\textrm{eV}^2]$ & $\sin^2\theta_{12}$ & $\sin^2\theta_{23}$ & $\sin^2\theta_{13}$ & $\delta$  \\
%\hline\hline
%3~$\sigma$  (NH)~&~ $7.11-8.18 $ ~&~ $2.30-2.65$ ~&~ $0.278-0.375$ ~&~ $0.392-0.643$ ~&~ $0.0177-0.0294$ ~&~ $0-2\pi$ ~ \\
%~\hskip 25pt(IH)           &~&~$2.20-2.54$           ~&~~&~$0.403-0.640$            ~&~$0.0183-0.0297$ ~&~\\
%\hline
%best-fit  (NH)   ~&~ $7.60$  ~&~ $2.48$  ~&~ $0.323$  ~&~ $0.567$  ~&~ $0.0234$  ~&~ $1.34 ~\pi$ \\
%\hline
%best-fit  (IH)  ~&~ $7.60$   ~&~ $2.38$  ~&~ $0.323$  ~&~ $0.573$  ~&~ $0.0240$  ~&~ $1.48 ~\pi$ \\
%\hline
%\end{tabular}
%\caption{ Allowed 3$\sigma$ ranges of oscillation parameters and best-fit values of these parameters
%for NH and IH \cite{Forero:2014bxa}.
% We have taken Majorana phase $\alpha$ to be zero in our analysis. 
%}
%\label{table:oscillation_param_2014}
%\end{table}

\section{\label{sec:nat}Constraints from Naturalness}

  %%%%%%%%%%%%% Begin OF FIGURE ################%%%%%%%%%%%%

\begin{figure}[t!]
\begin{center}
\includegraphics[width=10.0cm, angle
=0]{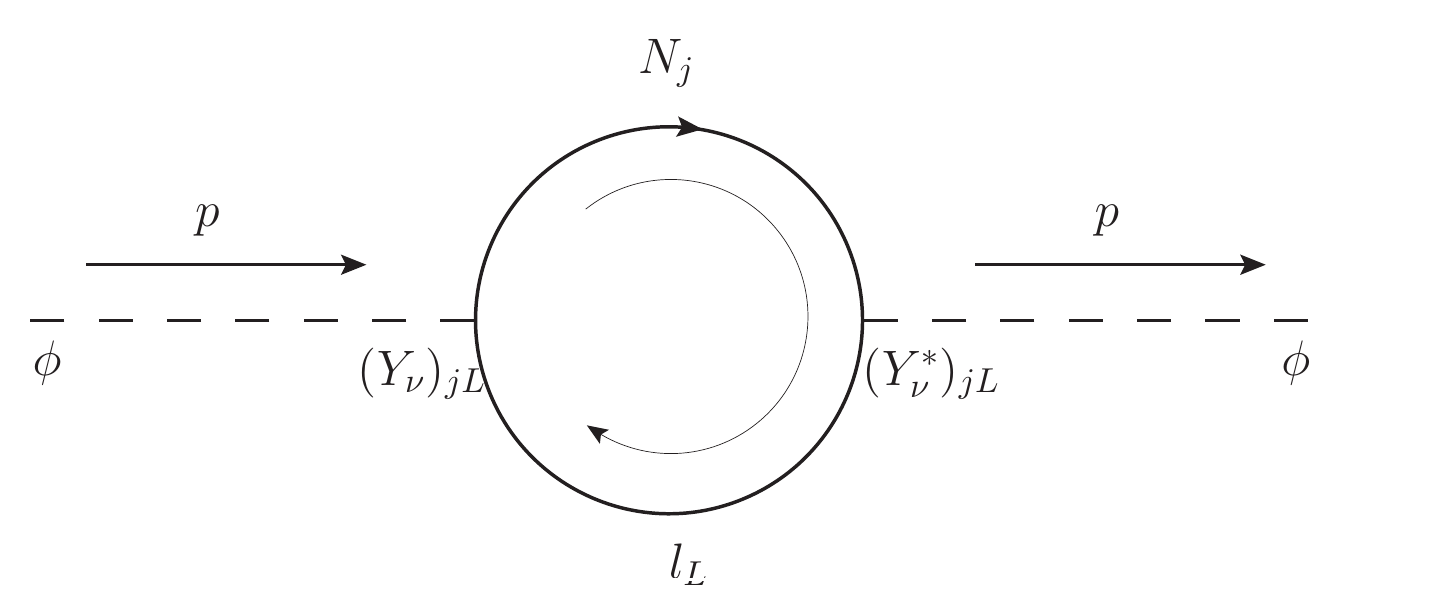}  
\end{center}
\caption{One-loop correction to the Higgs mass from neutrino Yukawa couplings due to the heavy neutrinos. The line inside the loop shows the orientation of lepton number. } 
 \label{fig:musquare_one_loop_correction_from_heavyneutrinos}  
\end{figure}
%%%%%%%%%%%%% End OF FIGURE %%%%%%%%%%%%%%%%%%%%%%%%%%%%%%%%%

We will discuss in this section the implications of the type-I seesaw from naturalness; see also Refs.~\cite{Vissani:1997ys,Clarke:2015gwa}. The potential of the SM Higgs boson at tree level can be written as
\begin{equation}
V \ = \ - \mu^2 (\phi^\dag \phi) + \lambda (\phi^\dag \phi)^2  \, ,
\end{equation}
from which the physical Higgs mass is given as
$m_h^2 = 2 \lambda v^2$. Heavy right-handed neutrino loop corrections to the electroweak $\mu$ parameter are desired to be smaller than ${\cal O}({\rm TeV}^2)$ for naturalness of the Higgs mass. 
The relevant diagram is shown in 
Fig.~\ref{fig:musquare_one_loop_correction_from_heavyneutrinos}. Using
the $\overline{\rm MS}$ scheme and taking the quantity
%\footnote{Note that there is a factor of
 % $\frac{1}{2}$ discrepancy in this quantity when compared to 
  %reference~\cite{Clarke:2015gwa}, where  the result is
  %$(\mbox{ln}[\frac{M_j}{\mu_R}] - \frac{1}{4} )$; the difference is
  %irrelevant for what follows. }
$(\mbox{ln}[\frac{M_j}{\mu_R}] - \frac{1}{2} )$ to be unity (where $\mu_R$
is the renormalization scale), the correction is given as 
\begin{equation}
 \delta{\mu^2} \ \approx \ \frac{1}{4\pi^2} {\rm Tr}[ Y_{\nu}^{\dagger}D_{N}^2Y_{\nu}]\, .
\end{equation}
Now using the CI parametrization from Eq.~(\ref{ynu}), we
get the simple relation
%\ggb{
\begin{equation}\label{eq_1}  
 \delta{\mu^2} \ \approx \ \frac{1}{4\pi^2} \frac{2}{v^2}{\rm Tr}[ {
   D_{\nu} R^{\dagger} D_{N}^3 R}] \ = \  
\frac{M_N^3 }{2\pi^2\,v^2} \cosh \left(2\, {\rm Im}[z]\right) \times \left\{ \begin{array}{cc}
(m_2 + m_3)   & \mbox{(NH)}\\
%    &=& 
 (m_1 + m_2)   & 
\mbox{(IH)} 
\end{array}\right.
    \end{equation}
Note that the real part of the complex angle $z$ in the CI
parametrization does not appear in Eq.~\eqref{eq_1}. The PMNS mixing angles
and CP phases also drop out in this expression. In the following discussion, the only relevant free parameters will therefore be $M_N$ and ${\rm Im}[z]$. Also note that for a given choice of ${\rm Im}[z]$, the correction in case of NH is about half the size of that in IH. 
%the term $\cos^2 z + \sin^2 z$ is in general different from $1$ for complex $z$ \blue{Not sure. Please check!}. Thus, 
%\begin{equation}\label{eq_2}  
% \delta{\mu^2} \approx 
%\left\{ \begin{array}{cc}
%\\
%  &=& 
%\frac{M_N^3}{2\pi^2\,v^2} ~ (m_2 + m_3) ~    & \mbox{(NH)}\\
%    &=& 
%\frac{M_N^3 }{2\pi^2\,v^2}  ~ (m_1 + m_2) ~\cosh \left(2\, {\rm Im}[z]\right)  & 
%\mbox{(IH)} 
%\end{array}\right.
%    \end{equation}
%\blue{So are you saying $\cos^2 z+\sin^2 z= \cosh (2{\rm Im}[z])$? A quick Mathematica check tells me it isn't correct. } 
%We see that
%If we choose $z$ to be real, then there is no dependence of
%$\delta{\mu^2}$ on $z$. 
%For complex $z$,   $\delta{\mu^2}$ can be
%significantly enhanced. 
%But if we choose $z$ to be complex, then  
%for both the cases mentioned above  
%$\delta{\mu^2}$ depends on $Im[z]$.  
%We have chosen $z$ to be purely imaginary, $z=  i Im[z]$.
%Also note that if $z$ is real, then
%${\rm Tr}({Y}_\nu^{\dagger}{Y}_\nu)$ is a very small number with no physical consequences. While in the case of complex or imaginary $z$, the Trace is enhanced by $\text{cosh}\left(2\,\text{Im}[z]\right)$.
%}
%We see that the real part of the complex angle $z$ in the Casas-Ibarra
%parametrization does not appear in this correction. The mixing angles
%and CP phases of the PMNS matrix also drop out. 

\begin{figure}[t]
\begin{center}
\includegraphics[width=7.5cm, angle=0]{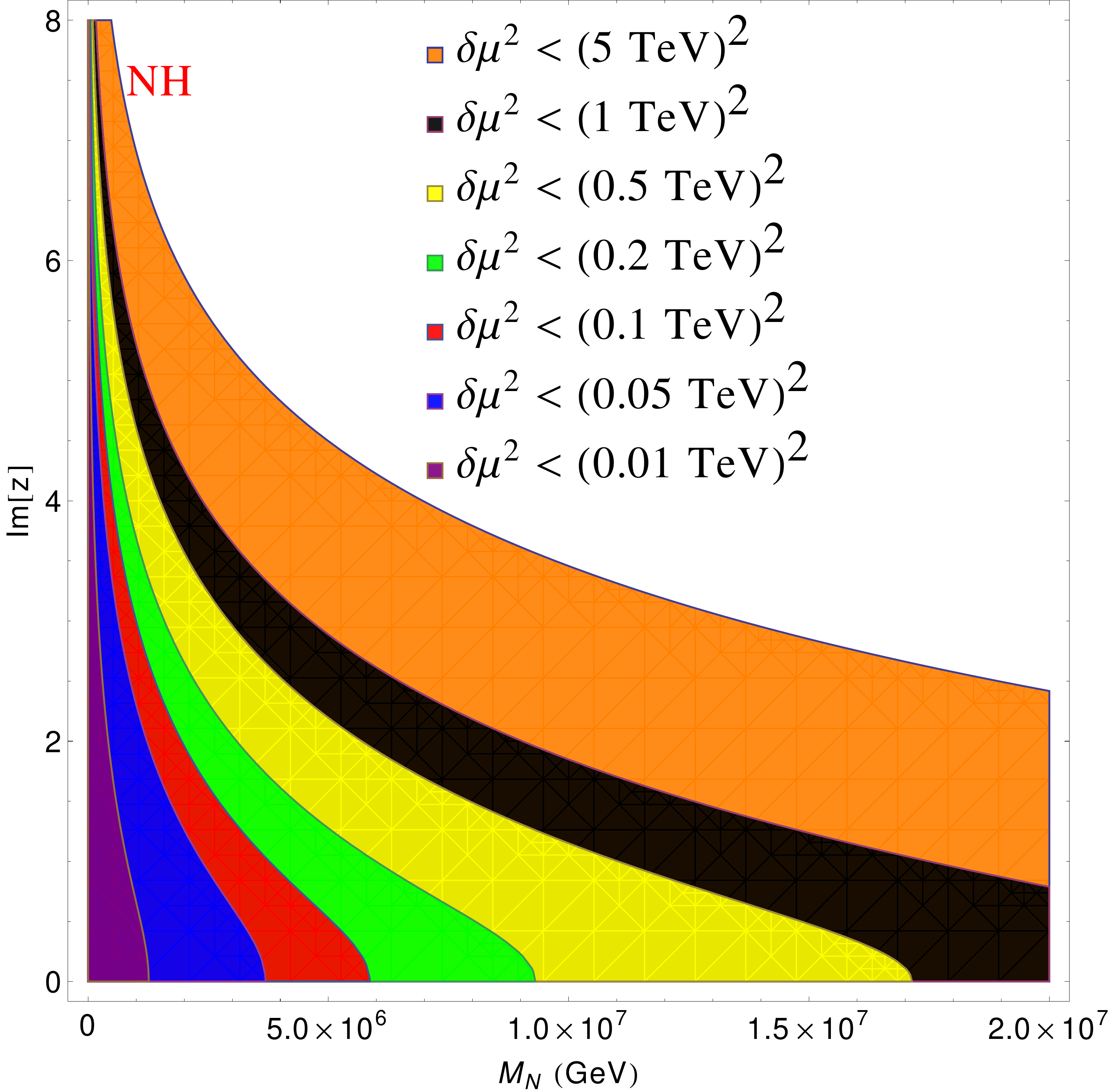} 
\includegraphics[width=7.5cm, angle=0]{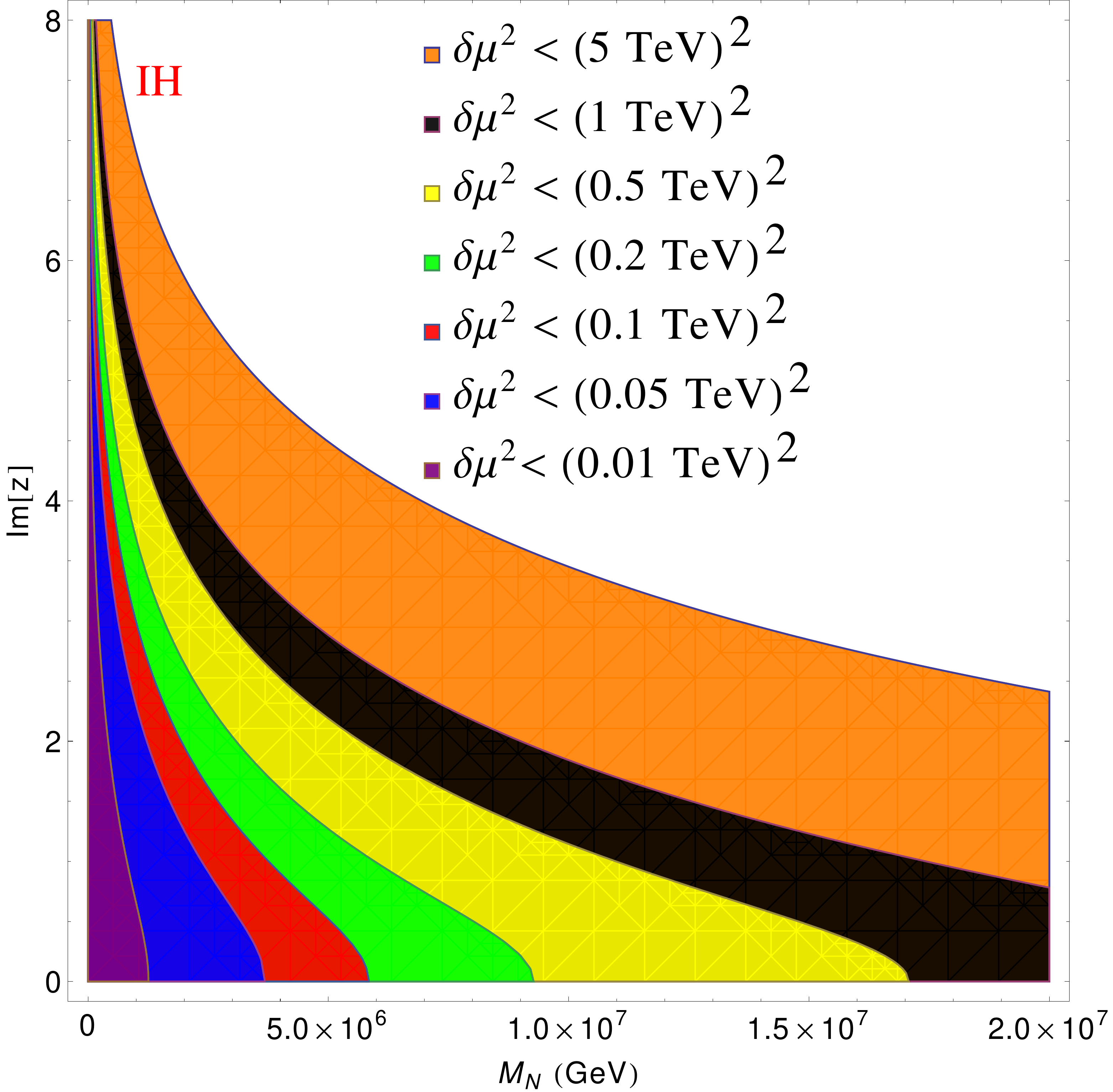} 
\caption{Naturalness contours in the $\text{Im}[z]$--$M_N$ plane. The left (right)
  plot is for NH (IH).  In the colored shaded regions, $\delta\mu^2$ is demanded to be below $(p\% ~\textrm{of 1 TeV})^2$, where $p = 500, 100,~50,~ 20,~ 10,~ 5,~ 1$ (from top to bottom). %From
%the plots we can see that as the mass of the heavy neutrino increases,
%the allowed values of Im$[z]$ decrease.
The unshaded regions are disfavored by naturalness. }
\label{fig:contourplot}
\end{center}
\end{figure}

In Figure~\ref{fig:contourplot} we show the regions in the Im$[z]$--$M_N$ plane 
corresponding to different upper limits of $\delta{\mu^2}$ ranging from $(5~\text{TeV})^2$ down to $(0.01 ~\text{TeV})^2$ for both NH and IH. 
The areas to the right of the shaded regions are thus disfavored from the 
condition of naturalness.  
%The $3\sigma$ ranges of the mass-squared differences 
%from reference \cite{Forero:2014bxa} have been used in obtaining these
%plots. 
From Figure~\ref{fig:contourplot}, we see that the larger Im$[z]$ is, the smaller the
allowed value of $M_N$ becomes. 
For instance, demanding $\delta \mu^2 < (1~{\rm TeV})^2$ implies $M_N < 2.7\times10^{7} ~\text{GeV}$ for Im$[z]=0$ and $M_N < 1.2\times10^{6} ~\text{GeV}$ for Im$[z]=5$. Similarly, demanding $\delta \mu^2 < (0.01~{\rm TeV})^2$ implies $M_N < 1.2\times10^{6}  ~\text{GeV}$ for Im$[z]=0$ and $M_N < 5.6\times10^{4} ~\text{GeV}$ for
Im$[z]=5$. In what follows, we will often use characteristic example
values, corresponding to the maximal values of  Im$[z]$ for a given 
$M_N$. These are, for $\delta \mu^2 < (1~{\rm TeV})^2$, Im$[z] = 8.75$ 
at $M_N = 10^5 ~\textrm{GeV}$ and Im$[z] = 11.17$ at $M_N =
2\times 10^4  ~\textrm{GeV}$. 

We should mention here that the naturalness constraints discussed above could in principle be relaxed if nature was supersymmetric at some scale below $\sim 10^7$ GeV. For instance, there are additional corrections to the Higgs mass from sneutrino loops which could in principle cancel those from RH neutrinos, if they have similar masses. However, we do not discuss this possibility here, simply because we are driven here by the minimality of the seesaw model.

\section{\label{sec:cons}Comparison with other bounds}
In this section we compare the constraints from naturalness obtained in Section~\ref{sec:nat} with those obtained from metastability of the electroweak vacuum (see Section~\ref{sec:vac}) and with phenomenological limits
arising from LFV (see Section~\ref{sec:lfv}). We also discuss implications for lepton number violating processes, such as $0\nu\beta\beta$ (see Section~\ref{sec:ndbd}). 
  
\subsection{\label{sec:vac}Bounds from Metastability}
% \textcolor{blue}{One needs to calculate  $Y_\nu^\dagger Y_\nu$ and take the trace 
% and check vacuum stability constraints. }
\begin{figure}[t]
\begin{center}
\includegraphics[width=7.7cm, height=6cm,angle=0]{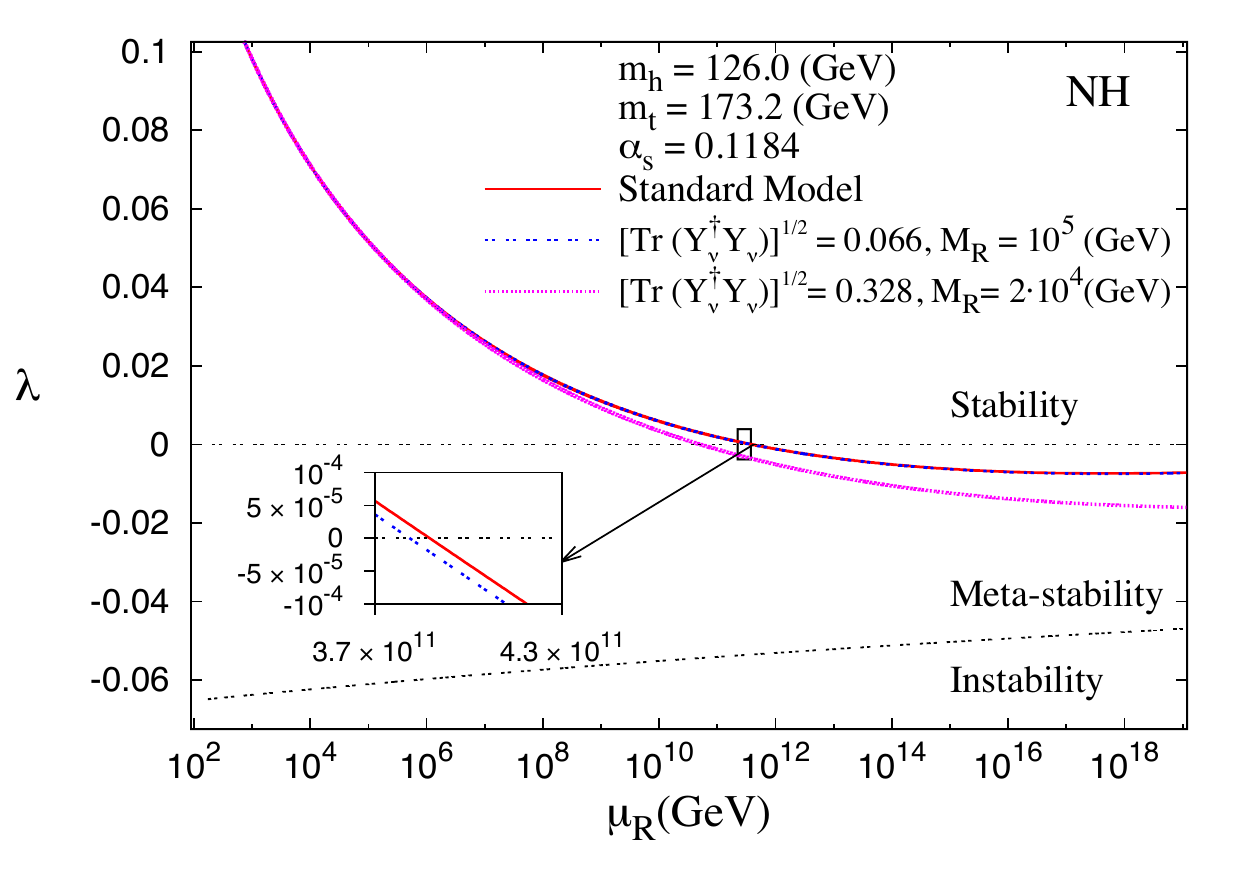} 
\includegraphics[width=7.7cm, height=6cm,angle=0]{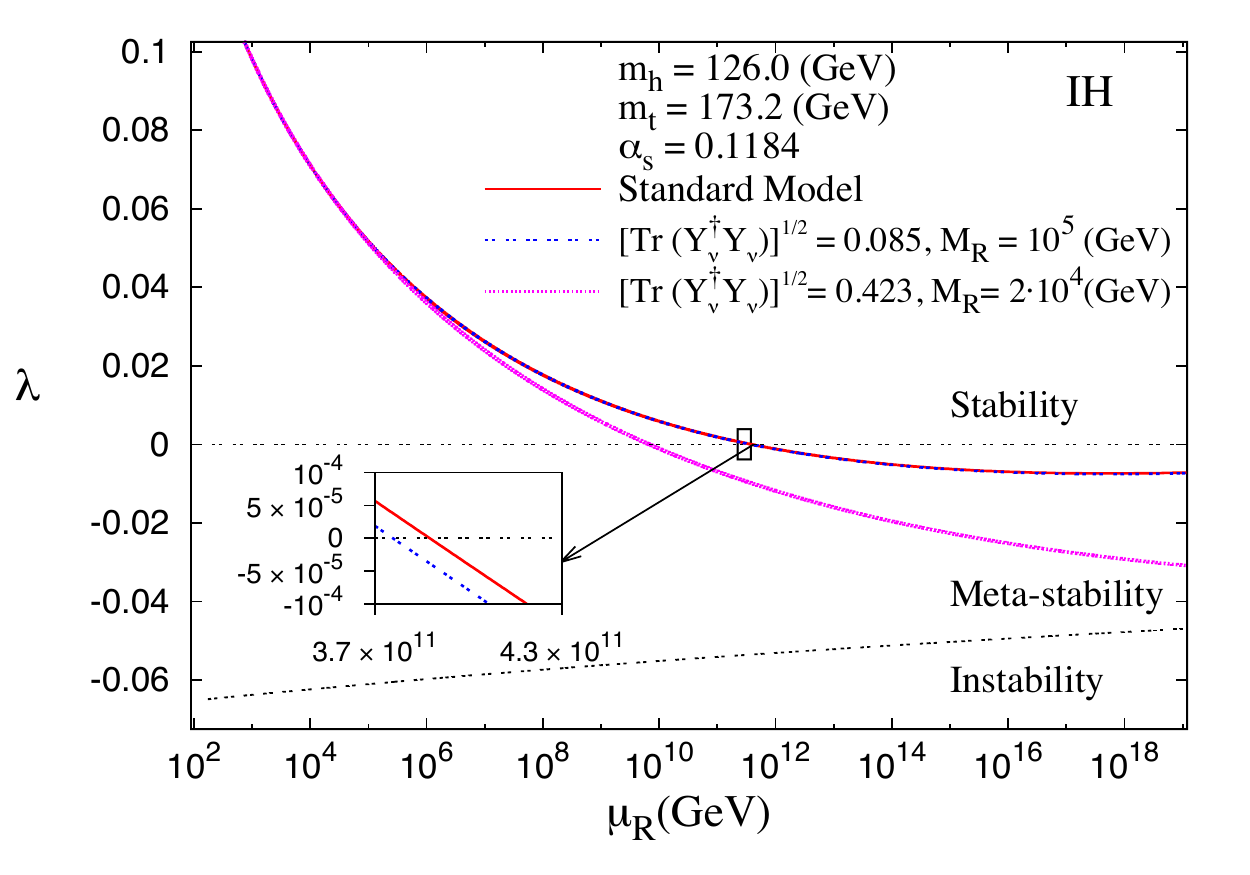} 
\caption{Running of the SM Higgs quartic coupling $\lambda$ for NH (left panel) and IH (right panel) in the minimal seesaw model. 
The red solid line corresponds to the SM running, while the blue
dotted line corresponds to the  SM + heavy neutrino contribution for $M_N = 10^5 ~\textrm{GeV}$.
For this mass  $\left(\text{Tr}[Y_\nu^{\dagger} Y_{\nu}]\right)^{1/2} = 0.066 $ is small and hence it almost overlaps with the SM line. 
In the inset both lines are shown with magnified view.
For $M_N = 2\times 10^4 ~\textrm{GeV}$ the maximal allowed
trace value is larger, $\left(\text{Tr}[Y_\nu^{\dagger}
  Y_{\nu}]\right)^{1/2} = 0.328 $, hence 
the effect is sizable and the corresponding pink line is visibly separated from the SM one. 
}
\label{fig:lam-vs-mu}
\end{center}
\end{figure}

We discuss in this section the constraints on the Yukawa coupling 
$Y_\nu$ and the heavy neutrino mass $M_N$ arising from 
metastability of the electroweak vacuum. The relevant RG equations can be found, e.g. in Refs.~\cite{Khan:2012zw, Lindner:2015qva} and are not repeated here. 
We just note here that the RG equations depend on the quantities  
\begin{align}
& {\rm Tr}({Y}_\nu^{\dagger}{Y}_\nu) %& =&  \frac{2}{v^2} {\rm Tr}({ [U
 %  D_{\sqrt{m_\nu}} R^{\dagger} D_{\sqrt{M_N}}] [ D_{\sqrt{M_N}} R
 %  D_{\sqrt{m_\nu}}U^\dagger]}) \nonumber \\
  \  = \ \frac{2}{v^2} {\rm Tr}\left[ { R^{\dagger} D_{N}  R D_{\nu}}\right]
  \ = \ \frac{2}{v^2} M_N \cosh\left(2\,\text{Im}[z]\right) \times 
\left\{\begin{array}{cc} 
 (m_2 + m_3) 
& \mbox{(NH)} \\
 (m_1 + m_2)
  &  \mbox{(IH)}
\end{array} \right. \label{eq:ynu2} \\
&{\rm Tr}({Y}_\nu^{\dagger}{Y}_\nu {Y}_\nu^{\dagger}{Y}_\nu) 
 \ = \ \frac{4}{v^4}  {\rm Tr}\left[( R^{\dagger} D_{N} R D_{\nu})^2\right]\nonumber\\
 & \ = \  \frac{4}{v^4} M_N^2 \times \left\{\begin{array}{cc}  
 \Big[(m_2^2 + m_3^2)\, \text{cosh}^2
 \left(2\,\text{Im}[z]\right) + 2 m_2 m_3\,
 \text{sinh}^2\left(2\,\text{Im}[z]\right) \Big]     & \mbox{(NH)} \\[0.3cm] 
     \Big[(m_1^2 + m_2^2)\, \text{cosh}^2
    \left(2\,\text{Im}[z]\right) + 2 m_1 m_2\,
    \text{sinh}^2\left(2\,\text{Im}[z]\right) \Big]  &  \mbox{(IH)} \\ 
%  \end{eqnarray}
% \right]
\end{array} \right. \label{eq:ynu4}
%\label{lambdatildaNR}
\end{align}
In Figure~\ref{fig:lam-vs-mu} we show the running of the quartic coupling 
$\lambda$ as a function of the renormalization scale $\mu_R$. 
We have chosen here and in the following the Higgs mass $m_h = 126 ~
\textrm{GeV}$, the top quark mass $m_t = 173.2 ~ \textrm{GeV}$ and the
strong coupling constant as $\alpha_s = 0.1184$ at the electroweak scale. 
The dashed horizontal line in Figure~\ref{fig:lam-vs-mu} shows the 
absolute stability condition for the 
vacuum, i.e.\ $\lambda > 0 $ for all $\mu_R$.  While  for sizable part of parameter 
space $\lambda < 0$ is reached,  it is however 
possible to find regions where the vacuum is metastable, 
i.e., the lifetime of the vacuum remains higher than the age of the Universe. 
Adopting a semi-classical approach, the tunneling probability at zero
temperature can be written as~\cite{Coleman:1977py,Callan:1977pt,Isidori:2001bm,Espinosa:2007qp}
\begin{eqnarray} \label{eq:mu}
p \ = \ \underset{\mu<\Lambda}{\text{max}}~V_U \, \mu^4 \,\,\text{exp}\left(-\frac{8\pi^2}{3|\lambda(\mu)|}\right) ,
\end{eqnarray} 
where $\Lambda$ denotes the cutoff scale and $V_U$ represents
the volume of the past light-cone which goes as $\tau^4$, with $\tau=4.35\times 10^{17}$ sec being the age of the
Universe~\cite{Ade:2015xua}. 
Metastability of the vacuum implies 
$p<1$, which in turn puts a lower bound on $\lambda$ as 
\begin{eqnarray} 
\left|\lambda\right| \ < \ \lambda_{\text{meta}}^{\text{max}} \ = \ \frac{8\pi^2}{3}\frac{1}{4\,\text{ln}\left(\tau\mu\right)} .
\label{lambda_meta}
\end{eqnarray}
We choose $\mu$ in Eq.~(\ref{eq:mu}) as the scale at which $\lambda$ becomes most negative.  %\blue{What is `most negative'?}
This constraint is shown by the slanting dashed line in Figure~\ref{fig:lam-vs-mu}. 

The red solid line in Figure~\ref{fig:lam-vs-mu} shows the running of ${{\lambda}}$
in the SM. 
The blue dashed line and the pink solid line show the running of 
this quantity in the minimal seesaw model considered here for two representative values of $M_N$. The blue dashed line is for $M_N = 10^{5}$ GeV for which the 
maximum value of Im$[z]$ allowed by naturalness is 8.75 [cf.\ Figure~\ref{fig:contourplot} and 
Eq.~\eqref{eq:ynu2}]. 
With this value of Im$[z]$, (Tr$[Y_\nu^\dagger Y_\nu])^{1/2}$ is found to  
be 0.085 and results only in a small difference with respect to the running in the
SM. 
In the inset we show a magnified version of  the region in $\mu_R$ where
the  stability limit is crossed for the SM and the minimal seesaw cases. It is clear from the 
inset that the stability is lost at a lower renormalization scale for the 
minimal seesaw model. 
The pink solid line shows the running of ${\lambda}$ for a lower value 
of $M_N = 2 \times 10^4$ GeV. This implies a higher maximum 
allowed value of Im$[z]=11.17$
for both NH and IH, and hence,  implies a larger (Tr$[Y_\nu^\dagger Y_\nu])^{1/2} = 0.423$.  
In this case the difference with respect to the SM running is clearly visible and the stability is lost earlier. However, in both cases the metastability bound remains 
satisfied. 

\begin{figure}[t]
\begin{center}
\includegraphics[width=7.7cm, height=6cm,angle=0]{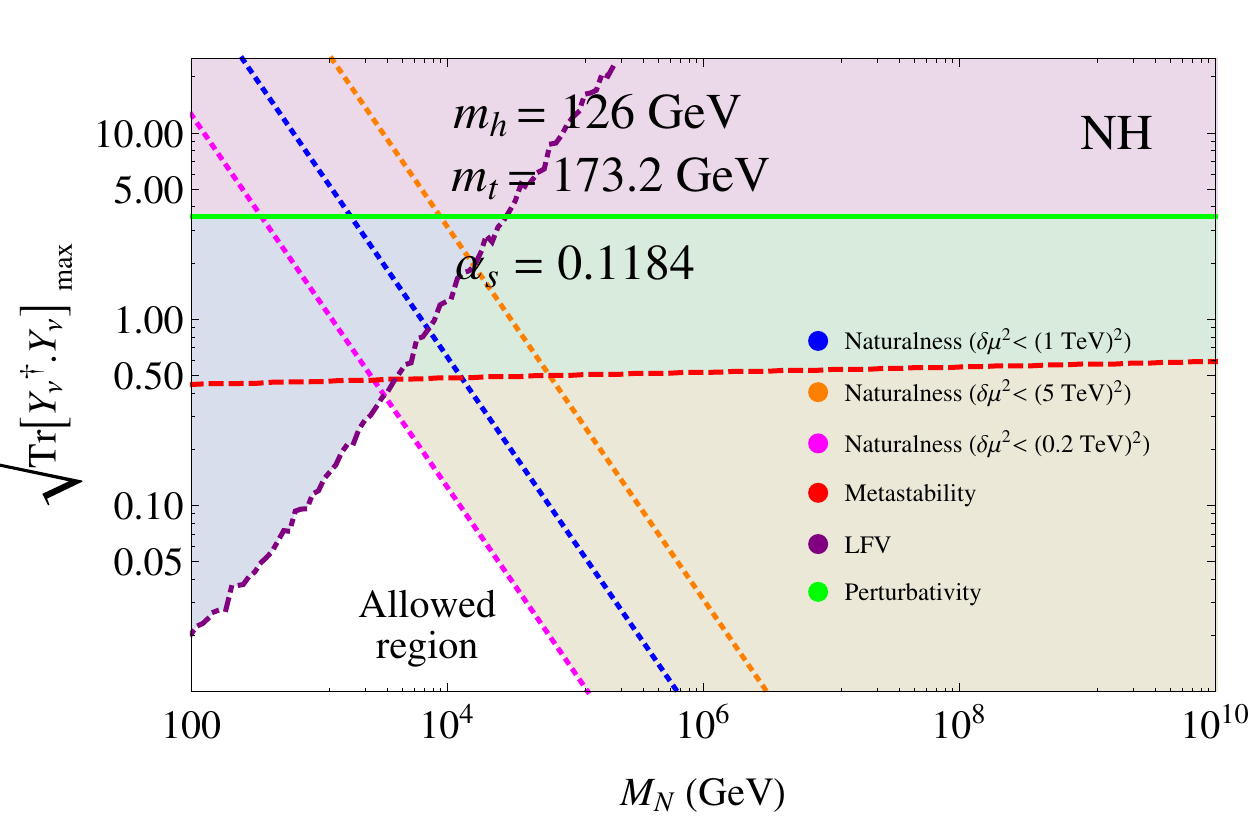} 
\includegraphics[width=7.7cm, height=6cm,angle=0]{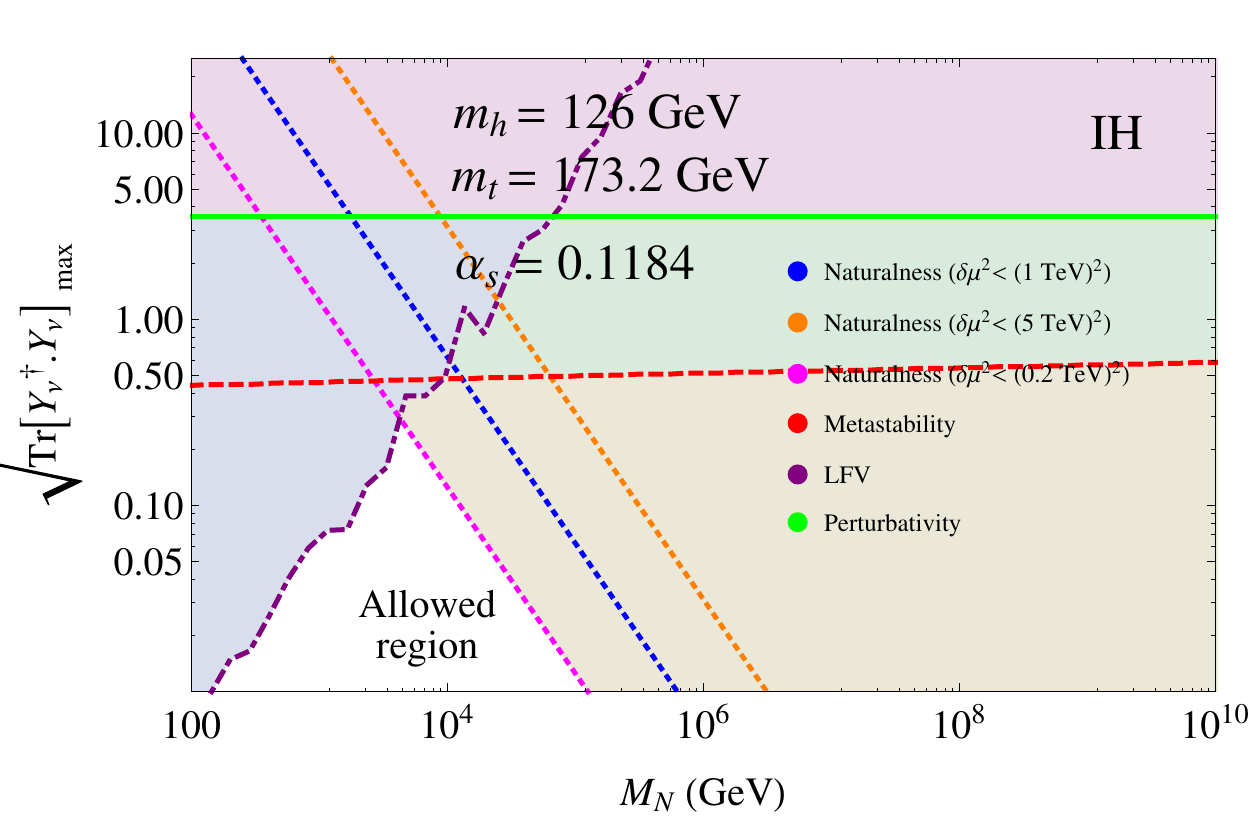} 
\caption{Bounds on $\left(\text{Tr}\left[Y_\nu^{\dagger} Y_{\nu}\right]\right)^{1/2}$ from metastability (red dashed), naturalness  (blue dotted), LFV (dark red dot-dashed) and perturbativity (green solid). 
%Perturbativity allows only the region below the green solid line, 
%while the metastability constraint allows only the region below the
%red dashed line. 
%When we impose the naturalness criterion $\delta \mu^2 < 1$ TeV$^2$, 
%the region above the blue dotted line is disfavored, while LFV
%constrains allows the region
%below the dark red  dot-dashed line. 
%Only the white region as allowed region by all these
%constraints. 
}
\label{fig:Cs_mediation}
\end{center}
\end{figure}

\begin{figure}[ht]
\begin{center}
\includegraphics[width=7.7cm, height=6cm,angle=0]{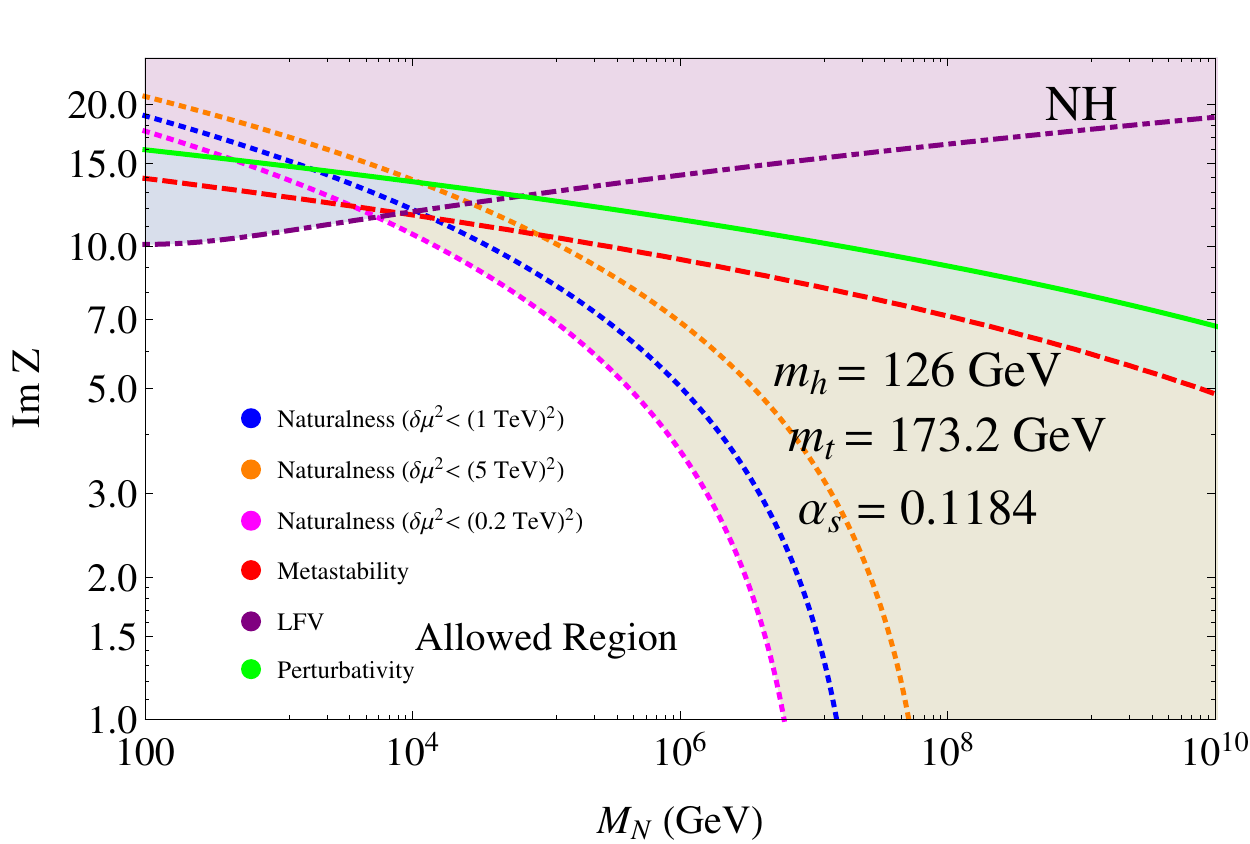} 
\includegraphics[width=7.7cm, height=6cm, angle=0]{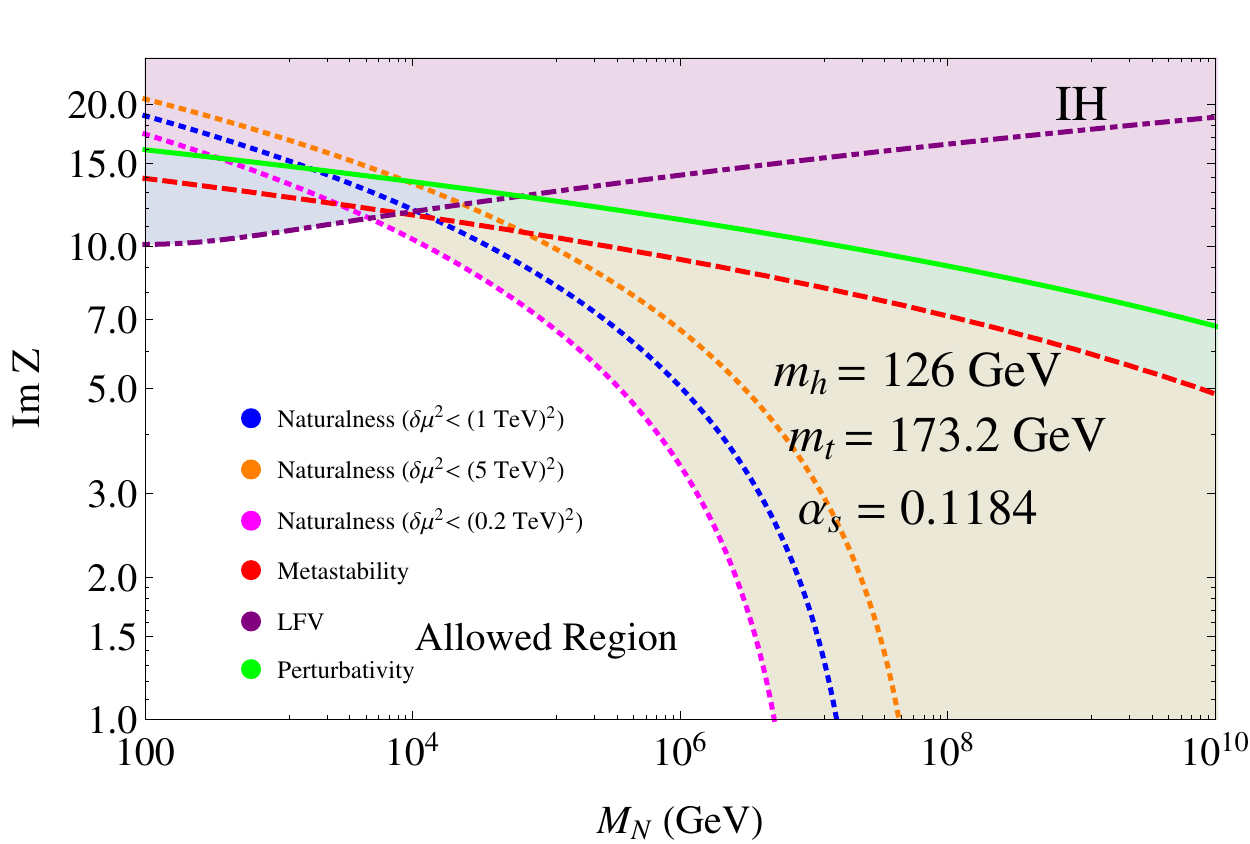} 
\caption{Bounds on Im$[z]$ as a function of $M_N$, with the same notation as in figure \ref{fig:Cs_mediation}. }
\label{fig:Cs_mediation1}
\end{center}
\end{figure}

The metastability condition can be used to impose 
an upper bound on $\text{Tr}[Y_\nu^\dag Y_\nu]$ from the running 
of $\lambda$ as a function of the heavy neutrino mass $M_N$. 
This is shown by the red-dashed line in Figure~\ref{fig:Cs_mediation} 
for both NH (left) and IH (right). 
The area below this curve is allowed from the metastability condition. 
This figure also shows the allowed region satisfying the condition of 
perturbativity ($\text{Tr}[Y_\nu^{\dagger}
    Y_{\nu}]\le 4\pi$), 
which is shown by the green solid line. 
The perturbativity bound is seen to be weaker than the metastability bound. 
For comparison, we also show the bound obtained from the naturalness criterion 
$\delta\mu^2 < (p~{\rm TeV})^2$ with $p=5,1,0.2$, as shown by the orange, blue and pink dotted lines  respectively, where the area to the left of these lines is preferred by naturalness. 
Figure~\ref{fig:Cs_mediation}  also contains bounds coming from
LFV considerations (see Section~\ref{sec:lfv}). 
Figure \ref{fig:Cs_mediation1} shows the same constraints in the
parameter space of Im$[z]$ and $M_N$.  

We find that for heavy neutrino masses larger than about $10^4$ GeV, naturalness
provides the strongest constraint on the minimal seesaw scenario. 
%\section{\label{sec:cons}Phenomenological Consequences} 


%\subsection{Other limits}
\subsection{Lepton Flavor Violation} \label{sec:lfv}
The strongest LFV bound on the minimal seesaw scenario comes from
the branching ratio for muon decay, $\mu \rightarrow e\gamma$, which is given
by \cite{Ilakovac:1994kj,Tommasini:1995ii,Dinh:2012bp}
\begin{eqnarray}
\text{Br} \left(\mu \rightarrow e\gamma\right)  \ = \  \frac{3\alpha_e}{ 8\pi}\left|T_{ei}T_{i\mu}^{\dag } f(x_i) \right|^2,
\label{eq:mu2egamma} 
\end{eqnarray}
where $\alpha_e\equiv e^2/4\pi$ is the fine-structure constant, $ x_i = \left(\frac{M_i^2}{m_W^2}\right)$ 
and 
\begin{eqnarray}
f(x) \ = \ \frac{x \left( 1-6x+3x^2+2x^3-6x^2 \ln x \right)} {2(1-x)^4}\,
\end{eqnarray}
is a slowly varying function of $x$ ranging from $0$ to $1$ for 
$x\in[0,\infty]$. 
The elements of the matrix $T$ denote the light-heavy mixing which 
is responsible for the non-unitarity of the lepton mixing matrix. 
In our scenario, $T = m_D^\dagger (M^{-1})^*U_R$ [cf. Eq.~\eqref{bdmatrix}], and $m_D$ is 
given in terms of the CI parametrization which introduces a
dependence on the PMNS matrix. The result for 
$\left|T_{ei}T_{i\mu}^{\dag } \right|^2$ is a lengthy expression
including the PMNS matrix elements, light and heavy neutrino masses,
as well as $z$, which is solved numerically. The resulting branching ratio has to be confronted with the latest experimental limit~\cite{TheMEG:2016wtm} of 
\begin{equation}
\text{Br} \left( \mu  \rightarrow  e\gamma\right ) \ < \ 4.2 \times 10^{-13} \,.
\label{brmu2e_MEG} 
\end{equation} 

The  LFV constraints 
on Im[$z$] and $M_N$, are displayed in Figure~\ref{fig:Cs_mediation1}.
In obtaining this, we 
vary the light neutrino mass and mixing parameters within their 
$3\sigma$ allowed range, the phases in the range $0-2\pi$ and take 
the extreme value that gives  maximum  disallowed region. 
The constraints on $z$ can be translated into constraints 
on Tr$([Y_\nu^\dagger Y_\nu])^{1/2}$ [cf.\ Eq.~\eqref{eq:ynu2}], which are shown 
in Figure~\ref{fig:Cs_mediation}. 
%That figure displays the limits on the scenario coming from naturalness, lepton flavor violation and metastability%\footnote{Constraints from 
%metastability and lepton flavor violation 
%in the context of different 
%seesaw scenarios were also studied in  
%\cite{Khan:2012zw,Bambhaniya:2014kga,Bambhaniya:2014hla}.}.     
%Combining the three constraints only  
%the white triangular region shown in Fig.~\ref{fig:Cs_mediation} 
%remains as the allowed region. 
Comparing with the naturalness, metastability and perturbativity constraints, we find that in the case of NH, LFV provides 
the strongest limits for relatively small heavy neutrino masses up to 
$M_N \sim 1$ TeV, then metastability takes over for a short mass
range, before naturalness imposes the strongest constraint on the minimal seesaw scenario. 
In the IH case, the situation is very similar, but for a given $M_N$, the constraint 
from LFV on the Yukawa couplings is slightly stronger in the IH case than in the 
NH case. This can be understood from the expression 
$x_{e\mu} \equiv \left|T_{ei}T_{i\mu}^{\dag } \right|^2$, which is larger for the 
IH case. In order to see it analytically, the lengthy formulas can be shortened in the limit of $\sin^2 \theta_{13}=0$,  $\sin^2 \theta_{23}=\frac 12 $ and 
$\sin^2 \theta_{12}=\frac 13$, and one finds 
\begin{align}
x_{e\mu}^{\rm NH}  & \ \approx \ \frac{2 \Delta m^2_{\rm a}}{M_N^2} 
\left( \cosh^2 (2 \,{\rm Im}[z]) \right) \, , \\
x_{e\mu}^{\rm IH} & \  \approx \ \frac{2 \Delta m^2_{\rm a}}{M_N^2} 
\left(\sinh^2 (2 \,{\rm Im}[z])  + \frac 19 \cosh^2 (2 \,{\rm Im}[z]) \right) \, ,
\end{align}
which illustrates that $x_{e\mu}$ is typically larger in the IH case for ${\rm Im}[z]\gtrsim 1$.

%Give details how this relates to  Tr$Y_\nu^\dagger Y_\nu$, as needed for Fig.4}
%below we give the values of these matrix for NH and IH. 
%{\bf should we put it in the appendix ?} 
%The current experimental limit from MEG collaboration \cite{Adam:2013mnn} is 
%\begin{equation}
%\text{Br} \left( \mu  \rightarrow  e\gamma\right ) < 5.7 \times 10^{-13} .
%\label{brmu2e_MEG} 
%\end{equation} 

%For masses around a few 100 GeV there are important constraints from 
%LFV, whereas for 
%$M_N = 10^4$ to $10^5 ~\textrm{GeV}$, the heavy neutrino contribution 
%to $\mu \rightarrow e\gamma$  will be negligible 
%as also evident from Fig.~\ref{fig:Cs_mediation} 
%\cite{Khan:2012zw,Bambhaniya:2014kga,Bambhaniya:2014hla}. 

\subsection{Neutrinoless double beta decay} \label{sec:ndbd}
% $0\nu\beta\beta$ section
%\subsection{$0\nu\beta\beta$}
Surprisingly, even for rather low values of the right-handed neutrino
masses, neutrinoless double beta decay $(0\nu\beta\beta)$~\cite{Rodejohann:2011mu} 
does not provide significant constraints in our scenario. 
The half-life for $0\nu\beta\beta$ in 
presence of heavy Majorana neutrinos is given by (see e.g. Refs.~\cite{Ibarra:2010xw, Tello:2010am, Mitra:2011qr, Chakrabortty:2012mh, Dev:2013vxa, Dev:2014xea})
\begin{eqnarray}
\frac{1}{T_{1/2}^{0\nu}} \ = \ 
{G}
\frac{|\mathcal{M}_{\nu }|^2}{m_e^2}
\left| U^2_{{ei}}\, m_i +
\langle p^2 \rangle \frac{T^2_{ei}}{M_i}
\right|^2,
\label{thalf} 
\end{eqnarray}
where $G$ denotes the phase space factor and $\langle p^2 \rangle = - m_e  m_p  \mathcal{M}_N/ \mathcal{M}_{\nu}$, 
whose magnitude is typically of order (100 MeV)$^2$. 
Here $ \mathcal{M}_{\nu} $ and $ \mathcal{M}_{N} $
denote the nuclear matrix elements corresponding to light and heavy neutrino
exchange, respectively. Using the general expression of 
$T= {m_D^{ \dagger}} (M^{-1})^{\ast} U_{R}$ from Eq.\ (\ref{bdmatrix}) 
and $Y_{\nu}$ from Eq.\ (\ref{ynu}), 
we can write 
$\sum_i T^2_{ei}/M_i= \sum_i U^2_{ei} m_i /M_N^2$, where $M_N$ is the
degenerate heavy neutrino mass. Substituting this in Eq.\ (\ref{thalf}), we find 
that the heavy-neutrino exchange contribution is suppressed by a factor of $\langle p^2 \rangle/M_N^2$, as compared to the light neutrino exchange contribution; see also Refs.~\cite{Xing:2009ce, Rodejohann:2009ve, Pavon:2015cga}.
% to   becomes 
%\begin{eqnarray}
%T_{(1/2)}^{-1}=
%{G}
%\frac{|\mathcal{M}_{\nu }|^2}{m_e^2}
%\left| U^2_{{ei}}\, m_i +
%\frac{\langle p^2 \rangle}{M_N^2} U^2_{{ei}}\, m_i
%\right|^2.
%\label{thalf2} 
%\end{eqnarray}
For heavy neutrino masses even as low as 100 GeV, the contribution of those
to the $0\nu\beta\beta$ half-life is therefore negligible. The contribution of 
light neutrinos is given by the usual expressions for a vanishing 
smallest neutrino mass, see e.g.\ Ref.\ \cite{Rodejohann:2011mu}. 
%The contribution of light neutrinos
%corresponds in the inverted hierarchy to half-lifes 
%Now, taking $\langle p^2 \rangle = (182 ~\textrm{MeV})^2 =  (182 \times 10^{-3} ~\textrm{GeV})^2$ and $M_N = 100 ~\textrm{GeV}$,  the contribution of heavy neutrinos compared that of the light neutrino
%is 
%$
%\frac{\langle p^2 \rangle}{M_N^2} = \frac{ (182 \times 10^{-3} ~\textrm{GeV})^2}{(100 ~\textrm{GeV})^2} = 3.3 \times 10^{-6}. 
%$
%It shows that even for heavy neutrino mass to be 100 GeV, the contribution of heavy neutrinos to $0\nu\beta\beta$ in comparison to light-neutrino contribution is negligible.
%}

%For $M_N = 10^4$-$10^5 \text{ GeV}$, the heavy neutrino contribution
%to $0\nu\beta\beta$ does not play a role in setting constraints. 

We should note here that electroweak-scale heavy neutrinos in the minimal seesaw can also be constrained from the LHC data using either the same-sign dilepton plus dijet channel (for scenarios with large lepton number violation) or opposite-sign dilepton or trilepton channels (for suppressed lepton number violation); for a review, see e.g.~\cite{Deppisch:2015qwa}. However, the current collider constraints turn out to be weaker than the other constraints shown in Figure~\ref{fig:Cs_mediation} for the range of heavy neutrino masses considered here.

\section{Running of Light Neutrino Parameters}\label{sec:rg}
In this section we discuss the RG evolution 
effect on light neutrino parameters 
in the context of the natural seesaw scenario considered in this work. 
We follow the procedure described in Ref.~\cite{Goswami:2013lba}. 
%The Lagrangian described by Eq.\ (\ref{lag:minimal_type1_seesaw})  
%gives rise to an effective dimension 5 operator below the 
%mass scale of the heavy fields: 
%\begin{equation}
% {\mathcal{L}_{eff}} = \frac{1}{4}\,\left(\overline{l^{{\cal{C}}}_{L}}\,\epsilon
% \, \phi\right)\kappa\left(\phi^T \epsilon^T l_{L}\right)+ h.c.,
%\end{equation}
%where $
%\begin{equation} 
%\kappa = 2 Y_\nu^T M_R^{-1} Y_\nu
%$.
%\label{seesaw1}
%\end{equation}
%The mass matrix, $m_\nu$ is obtained as $m_\nu = \frac 14 \kappa \, v^2 $ 
%and can be diagonalized by 
%\begin{eqnarray}
%$
%U^T m_{\nu} U = m_{\rm diag}
%$,
%\end{eqnarray}
%
%where $U$ is the  PMNS matrix in the basis where the 
%charged lepton matrices are diagonal.
%

\begin{figure}[t!]
\begin{center}
\includegraphics[width=7.7cm, angle=0]{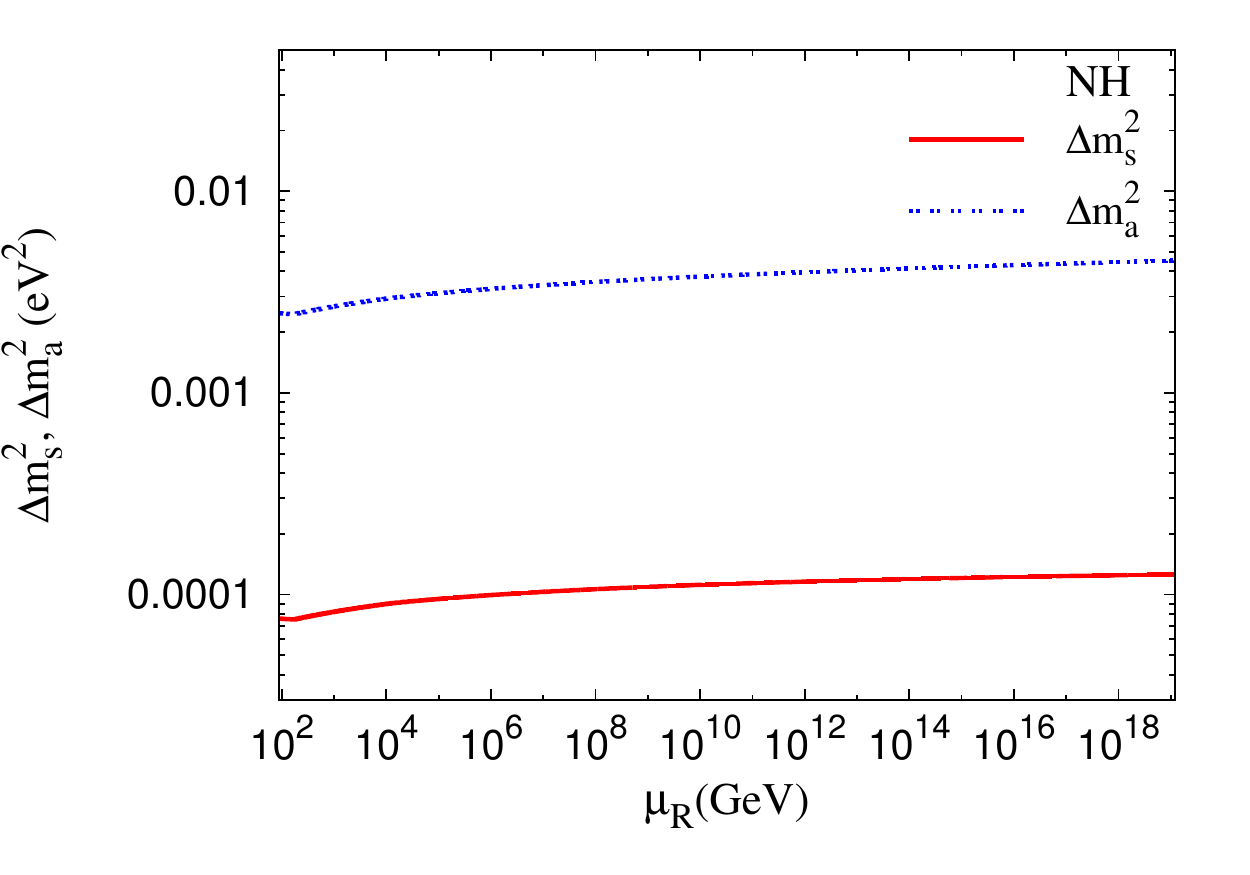} 
\includegraphics[width=7.7cm, angle=0]{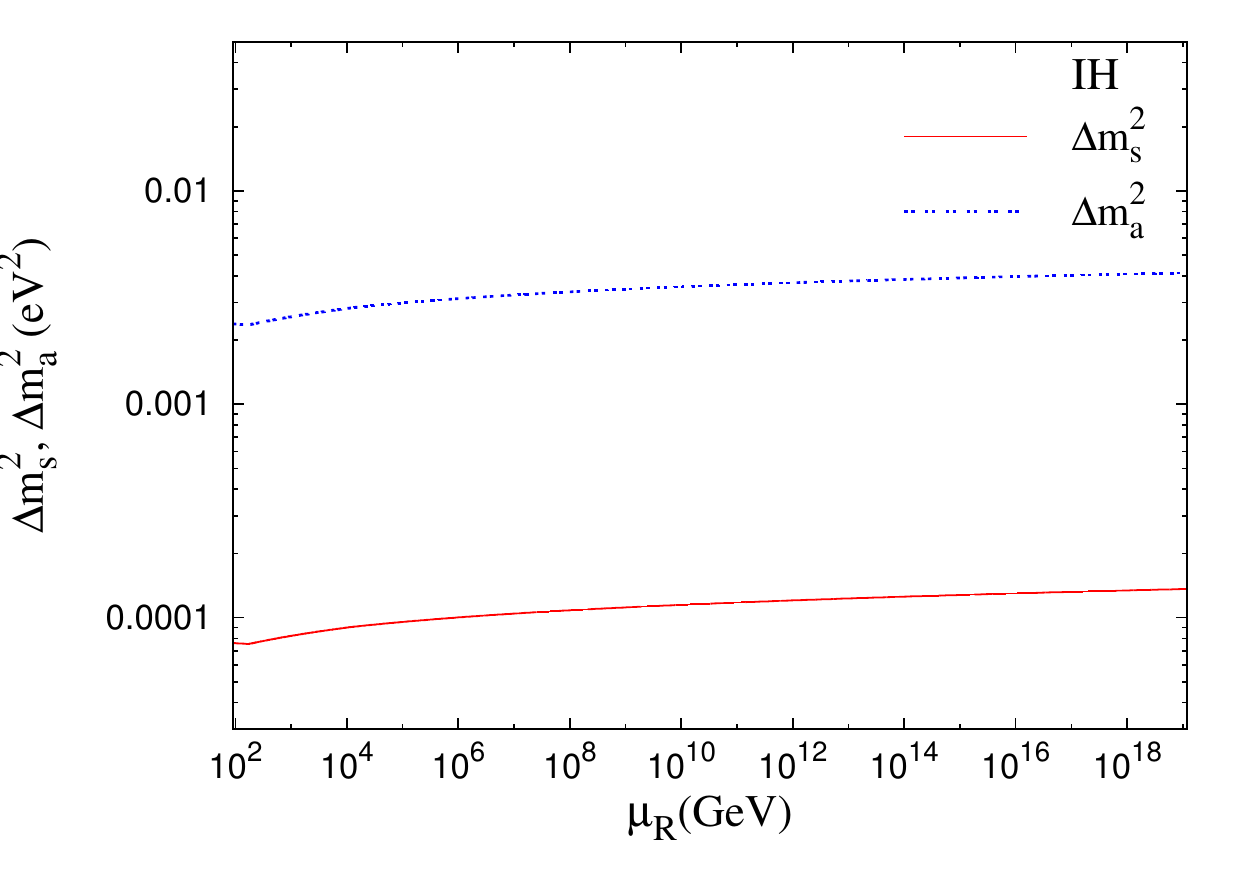} 
\caption{Running of solar and atmospheric mass-squared differences
  with the renormalization scale. The plots correspond 
to the central values of the oscillation parameters, the Majorana 
phase is chosen to be $\pi/4$.}
\label{fig:Delta_mass-sq_vs_mu}
\end{center}
\end{figure}
\begin{figure}[t!]
\begin{center}
\includegraphics[width=7.7cm, angle=0]{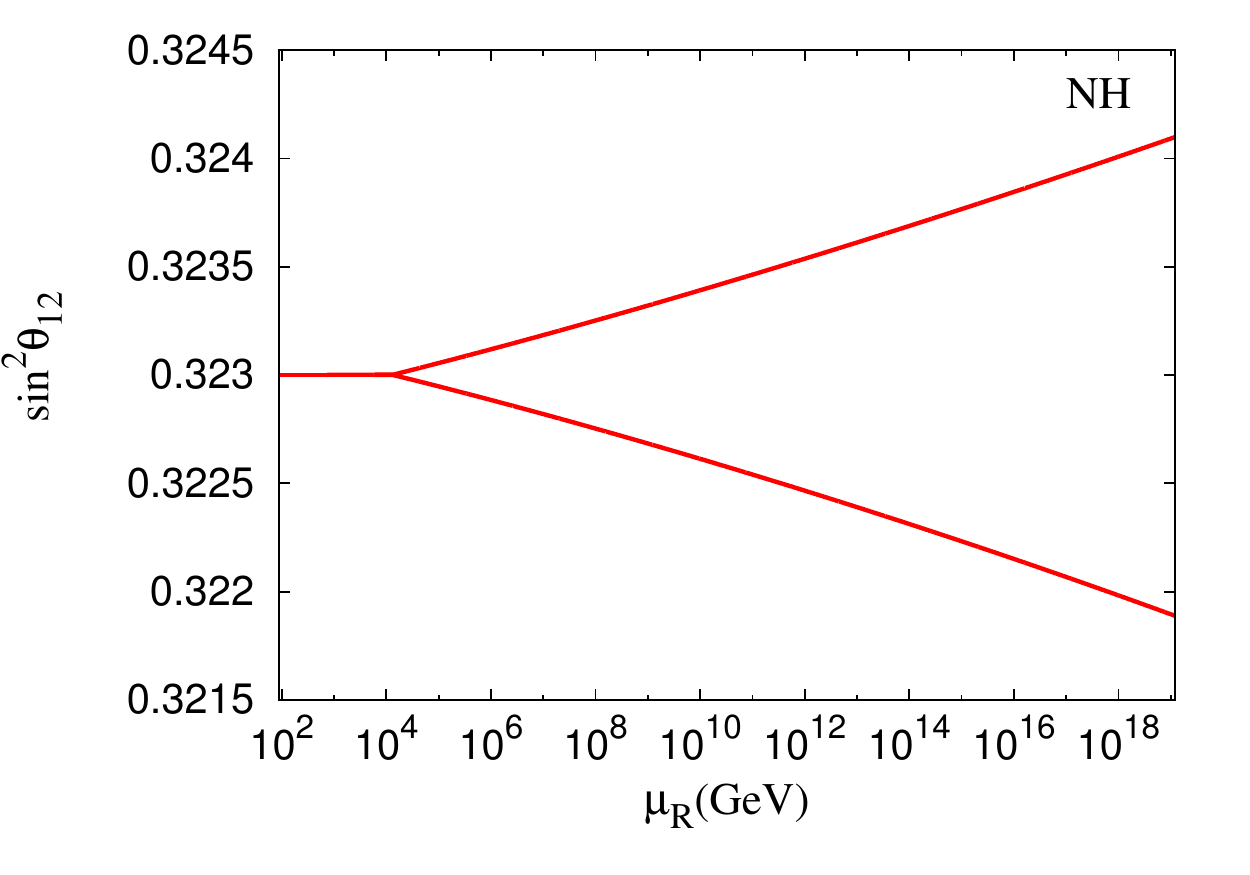} 
\includegraphics[width=7.7cm, angle=0]{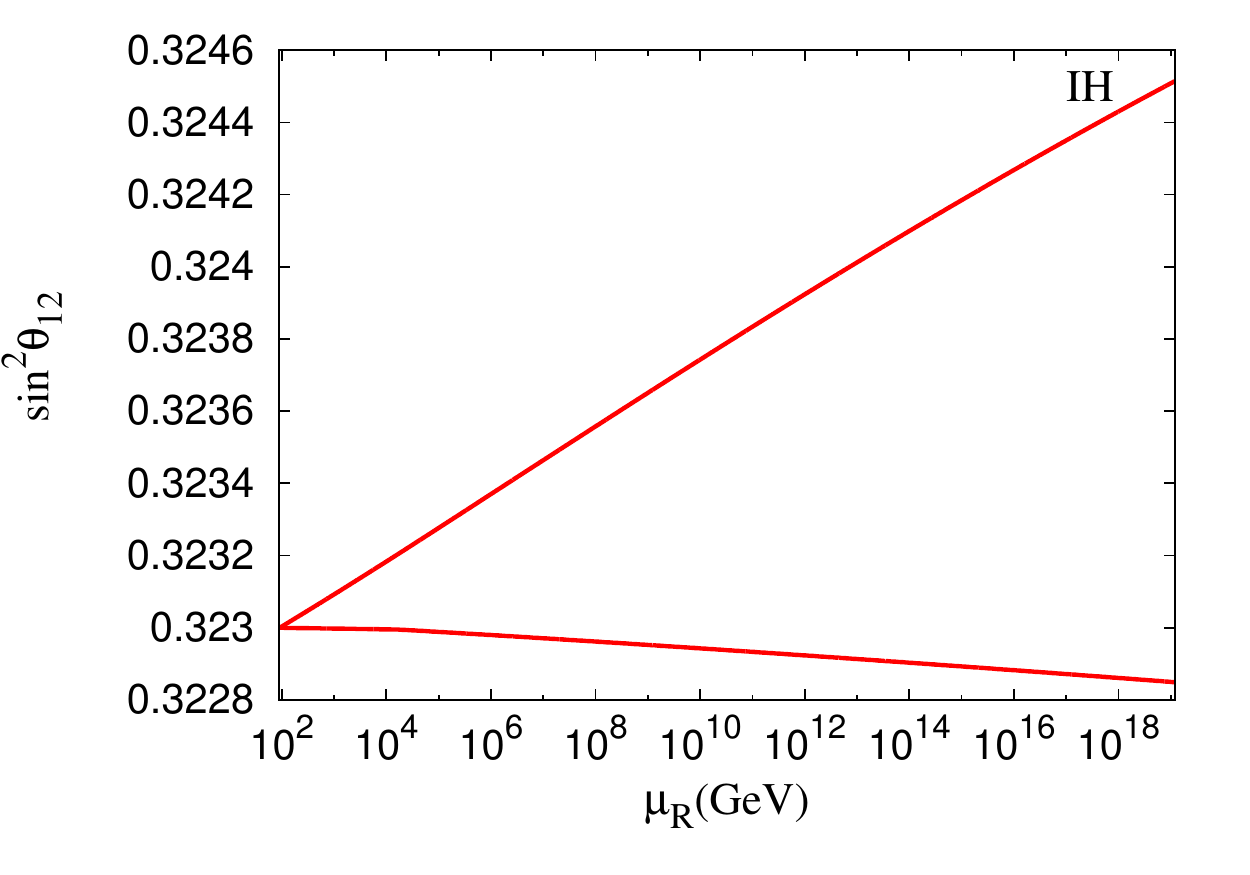} 
\caption{Running of $\sin^2 \theta_{12}$ with the renormalization
  scale for NH and IH. The plots correspond 
to the central value of $\sin^2 \theta_{12}$, all other parameters
 are varied in their $3\sigma$ ranges.}
\label{fig:angles12-nh_vs_mu}
\end{center}
\end{figure}

Starting with low energy parameters, the running of light 
neutrino parameters \cite{Antusch:2003kp} 
is governed by the effective dimension-5 operator 
\begin{equation}
 {\mathcal{L}_{\rm eff}} \ = \ \frac{1}{4}\,\left(\overline{l^{{\cal{C}}}_{L}}\,\epsilon
 \, \phi\right)\kappa\left(\phi^T \epsilon^T l_{L}\right)+ {\rm H.c.} ,
\end{equation}
where $\kappa = 2 Y_\nu^T M_R^{-1} Y_\nu = 4 \, m_\nu/v^2$. When the energy 
scale of the heavy neutrino masses is reached, different RG equations of the 
full renormalizable theory have to be considered~\cite{Antusch:2005gp}, and threshold 
effects can be important. We solve the evolution equations  
numerically taking into account the threshold effect at the mass scale 
of the degenerate neutrinos. 
In our analysis for this subsection, we take $M_N = 1.32\times 10^4$ 
GeV for which the maximum allowed value of 
$\text{Im}[z]$ is $11.797~(11.544)$ for
NH (IH).  

%$\left(\text{Tr}\left[Y_\nu^{\dagger} Y_{\nu}\right]\right)^{1/2} = 0.476$. 
%{\bf{subjected to or correspond to ?}} 

Figure \ref{fig:Delta_mass-sq_vs_mu} shows the running of the 
solar 
and atmospheric mass-squared differences $\Delta m^2_s \equiv  m_2^2 -m_1^2$  
and $\Delta m^2_{a} \equiv m_3^2 - m_2^2$ for NH and $m_1^2 - m_3^2 $ for IH.  
The figures show that the mass parameters do not run much. 
This is expected since the running of the masses is proportional 
to the masses themselves and it is well known that 
for hierarchical neutrinos the running is not very significant \cite{Antusch:2003kp}.

\begin{figure}[t!]
\begin{center}
\includegraphics[width=7.7cm, angle=0]{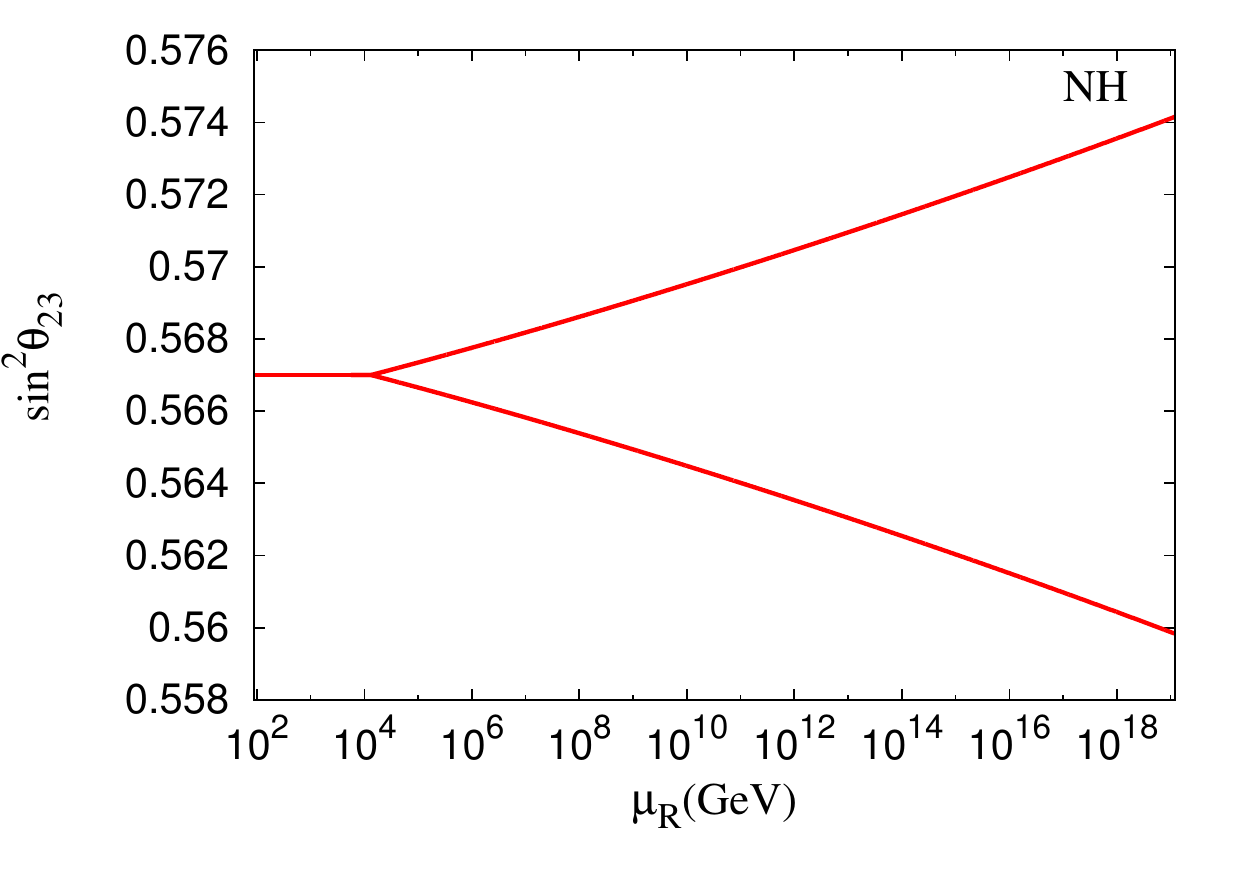}
\includegraphics[width=7.7cm, angle=0]{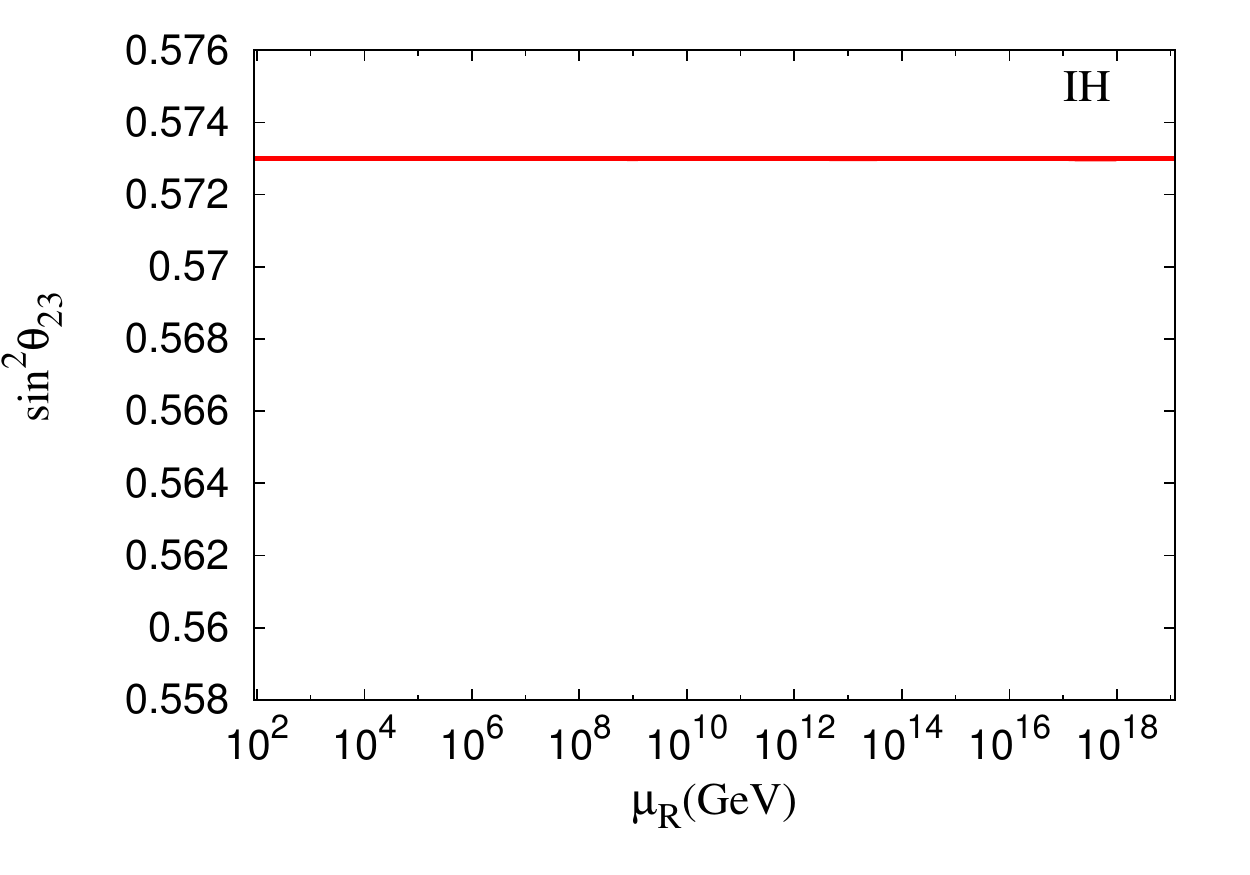}
\caption{Running of $\sin^2 \theta_{23}$ with the renormalization
  scale for NH and IH. The plots correspond 
to the central value of $\sin^2 \theta_{23}$, all other parameters
 are varied in their $3\sigma$ ranges.}
\label{fig:angles23-nh_vs_mu}
\end{center}
\end{figure}
\begin{figure}[t!]
\begin{center}
\includegraphics[width=7.7cm, angle=0]{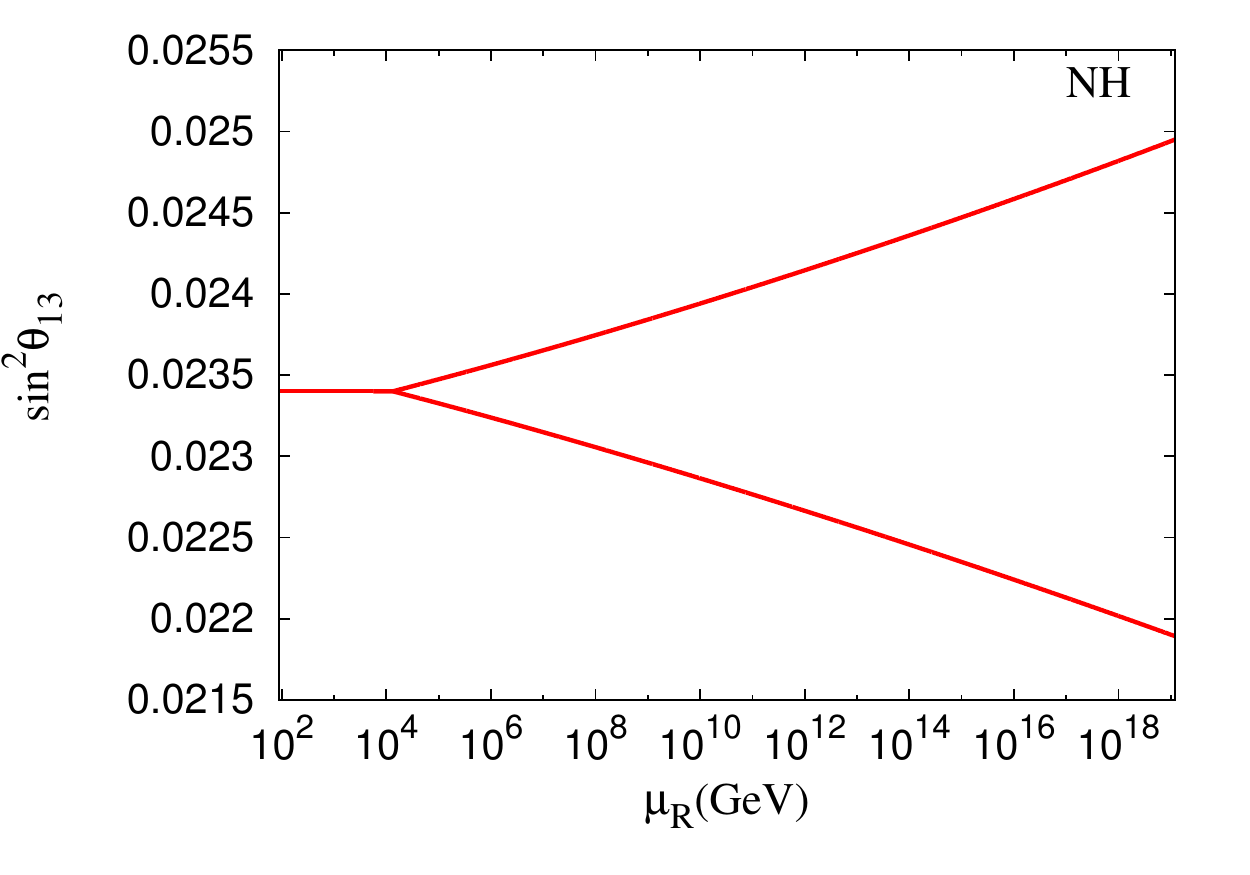} 
\includegraphics[width=7.7cm, angle=0]{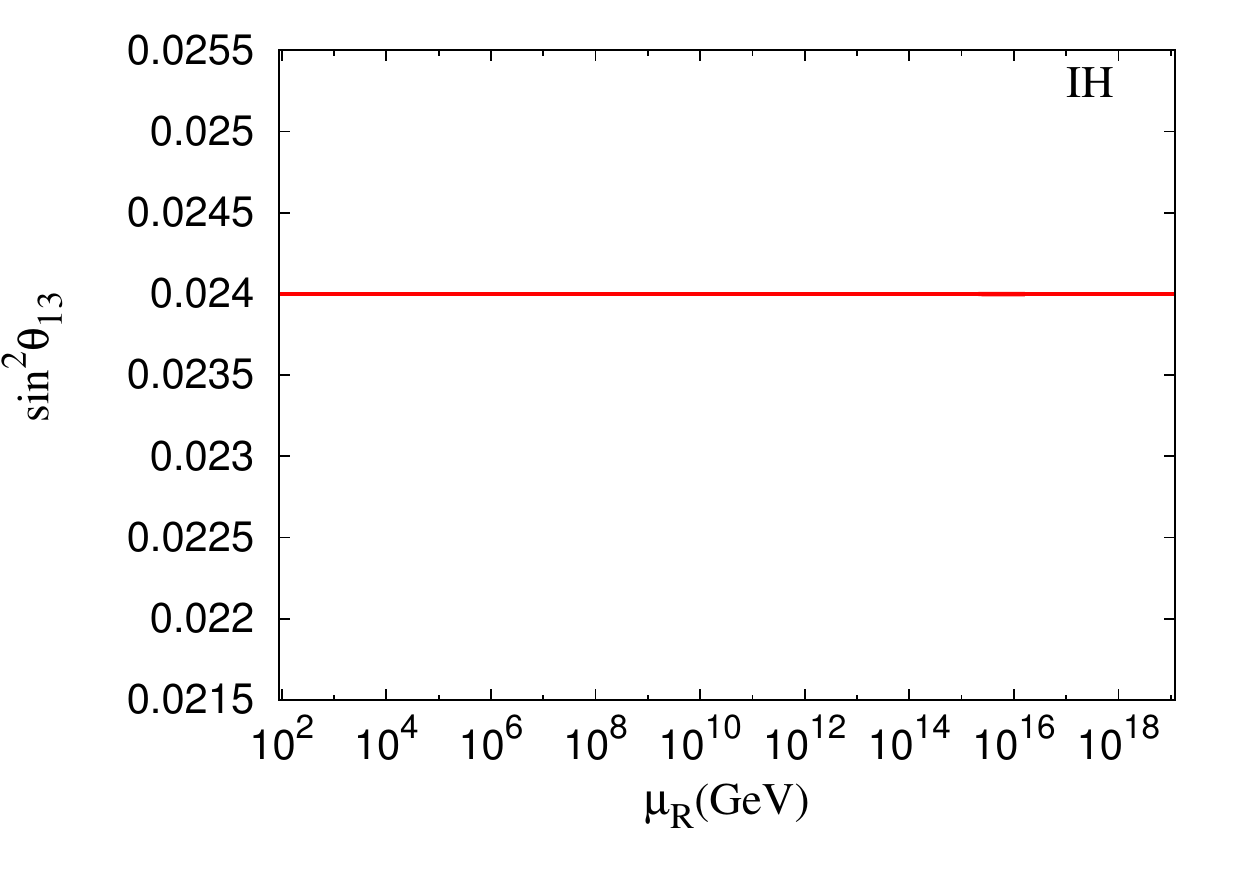} 
\caption{Running of $\sin^2 \theta_{13}$ with the renormalization
  scale for NH and IH. The plots correspond 
to the central value of $\sin^2 \theta_{13}$, all other parameters
 are varied in their $3\sigma$ ranges.}
\label{fig:angles13-nh_vs_mu}
\end{center}
\end{figure}

Figure \ref{fig:angles12-nh_vs_mu} shows the running of the 
mixing angle $\sin^2\theta_{12}$ for NH and IH.  
Note that due to the presence of threshold effects the 
running is not unidirectional and  while running from low to high scale the 
value of the mixing angle can either increase or decrease 
\cite{Goswami:2013lba}. 
The figure shows the maximum running of this angle  in both directions 
by varying all the oscillation parameters in their allowed $3\sigma$
range and both $CP$ phases in the range $-\pi$ to $+\pi$.   
It is seen that although the high-scale value can be lower or higher than
the low scale value the dispersion   due to RG effects is not large.

Figure \ref{fig:angles23-nh_vs_mu} shows the running of the 
mixing angle $\sin^2\theta_{23}$ for NH and IH.   
We take the low scale value to be in the higher octant, 
$\theta_{23} > \pi/4$. 
The running for NH can 
be in both directions due to the threshold effect and we show the 
maximum and minimum amount of running obtained by varying the parameters
in their  $3\sigma$ ranges. Note that even after running, the 
octant does not change between low and high scale.  
For IH,
$\sin^2 \theta_{23}$ does not run as the running is proportional to $m_3$ 
which is zero in our model. 
This is also true for $\sin^2\theta_{13}$ and this is reflected in 
the right plot of figure~\ref{fig:angles13-nh_vs_mu} which shows the 
running of $\sin^2\theta_{13}$ for IH. 
For NH, again due to threshold effects the high scale value can be lower or 
higher than the low-scale value. 

%\blue{What is the connection of this analysis to the previous sections?} 

%\begin{figure}[htb]
%\begin{center}
%\includegraphics[width=7.7cm, angle=0]{alpha-delta-vs-mu_nh.eps} 
%\includegraphics[width=7.7cm, angle=0]{alpha-delta-vs-mu_ih.eps} 
%\caption{Running of the Dirac and Majorana CP phase with the
%  renormalization scale. The plots correspond 
%to the central value of all oscillation parameters, the Majorana phase is chosen to be $\pi/4$.}
%\label{fig:alp-del_vs_mu}
%\end{center}
%\end{figure}

%Finally the figure \ref{fig:alp-del_vs_mu} give the running of the 
%Dirac and Majorana phases.  

%{\bf details, discussion of plot results, why bifurcation? Why no
%  bifurcation for theta13 and theta23 in IH case? Why Majorana phase
%  not running? Is alpha=pi/4 fixed point?}

The RG running discussed above will be somewhat modified for the non-degenerate and/or 3 RH neutrino case.  However, we expect our results to be valid to, say 0.005 eV for the smallest neutrino mass, beyond which typical enhancement factors of RG running as in Table 2 of Ref.~\cite{Antusch:2003kp} will apply. Making the RH neutrinos non-degenerate makes the RG running essentially unpredictable unless one specifies exactly their masses. This also takes us beyond the minimality, which is the main topic of this paper.

\section{\label{sec:yb}Leptogenesis}
A cosmological consequence of the type-I seesaw mechanism is that the $CP$-violating decays of the heavy Majorana neutrinos can explain the observed BAU by the mechanism of leptogenesis~\cite{Fukugita:1986hr}. However, the naturalness and other constraints discussed in Section~\ref{sec:cons} (see Figure~\ref{fig:Cs_mediation1}) disfavor 
\cite{Clarke:2015gwa} standard thermal leptogenesis, which requires $M_N\gtrsim 10^9$ GeV~\cite{Davidson:2002qv, Giudice:2003jh}. As we will show in this section, resonant leptogenesis (RL)~\cite{Flanz:1996fb, Pilaftsis:1997jf, Pilaftsis:2003gt} is a viable alternative to explain the observed BAU in the minimal natural seesaw. 

In the RL mechanism, a small mass splitting $\Delta M_N$ between the two heavy neutrinos of the order of their average decay width $\Gamma_N$ leads to the resonant enhancement of the $\varepsilon$-type $CP$ asymmetry~\cite{Pilaftsis:1997jf}. A minimal way to motivate the quasi-degeneracy between the heavy neutrinos is by radiative effects~\cite{GonzalezFelipe:2003fi, Turzynski:2004xy, Branco:2005ye}, starting from a degenerate spectrum at some high scale~\cite{Pilaftsis:2005rv, Deppisch:2010fr, Dev:2014laa}. However, this minimal scenario is not viable here, as the leptonic $CP$ asymmetry vanishes identically at ${\cal O}(Y_\nu^4)$~\cite{Dev:2015wpa}. To avoid this no-go theorem, one needs to include a new source of flavor breaking mass splitting in the heavy-neutrino sector, which could in principle be motivated from the approximate breaking of some flavor symmetry in the leptonic sector. Here we simply assume this to be the case and choose the mass splitting $\Delta M_N \simeq \Gamma_N/2$ to maximize the $CP$ asymmetry, without investigating the details of how this could be generated in a concrete flavor model. The only relevant effect of this small splitting is within leptogenesis. 

The flavored $CP$ asymmetry due to heavy-neutrino mixing is given by (see e.g. Ref.~\cite{Dev:2014laa}) 
\begin{align}
\varepsilon_{i l}^{\rm mix} \ = \  \sum_{j\neq i} \frac{{\rm Im}\left[Y_{\nu_{i l}}Y^*_{\nu_{j l}}(Y_\nu Y^\dag_\nu)_{ij} \right]+\frac{M_{i}}{M_{j}}{\rm Im}\left[Y_{\nu_{i l}}Y^*_{\nu_{j l}}(Y_\nu Y^\dag_\nu)_{ji} \right]}{(Y_\nu Y^\dag_\nu)_{ii} (Y_\nu Y^\dag_\nu)_{jj}} f_{ij}^{\rm mix} \, ,
\label{eq:mix}
\end{align}
with the regulator given by 
\begin{align}
f_{ij}^{\rm mix} \ = \ \frac{(M_{i}^2-M_{j}^2)M_{i}\Gamma_{j}}{(M_{i}^2-M_{j}^2)^2+M_{i}^2\Gamma_{j}^2} \, ,
\end{align}
where $\Gamma_{i}=(M_{i}/8\pi)(Y_\nu Y_\nu^\dag)_{ii}$ is the tree-level heavy-neutrino decay width. 
There is a similar contribution $\varepsilon_{i l}^{\rm osc}$ to the $CP$ asymmetry from heavy-neutrino oscillations~\cite{Dev:2014laa, Dev:2014wsa}, which is formally at ${\cal O}(Y_\nu^4)$, different from the ${\cal O}(Y_\nu^6)$ effect for GeV-scale seesaw as considered in Refs.~\cite{Akhmedov:1998qx, Canetti:2012kh, Shuve:2014zua}. Its form is given by Eq.~\eqref{eq:mix} with the replacement $f_{ij}^{\rm mix}\to f_{ij}^{\rm osc}$, where~\cite{Dev:2014laa, Dev:2014wsa}
\begin{align}
f_{ij}^{\rm osc} \ = \ \frac{(M_{i}^2-M_{j}^2)M_i\Gamma_{j}}{(M_{i}^2-M_{j}^2)^2+(M_i\Gamma_{i}+M_j\Gamma_{j})^2\frac{{\rm det}\left[{\rm Re}(Y_\nu Y^\dag_\nu)\right]}{(Y_\nu Y^\dag_\nu)_{ii} (Y_\nu Y^\dag_\nu)_{jj}}} \, .
\end{align}
The total $CP$ asymmetry is thus given by $\varepsilon_{i l}=\varepsilon_{i l}^{\rm mix}+\varepsilon_{i l}^{\rm osc}$. 
 
After solving the relevant flavored Boltzmann equations and taking into account the appropriate efficiency and dilution factors (for details, see e.g. Refs.~\cite{Buchmuller:2004nz, Deppisch:2010fr, Dev:2014laa}), the final BAU $\eta_B\equiv n_B/n_\gamma$ (where $n_B, n_\gamma$ are the number densities of baryons and photons today) can be written analytically as 
\begin{align}
\eta_B \ \simeq \ -\frac{28}{51}\frac{1}{27}\frac{3}{2} \sum_{l,i} \frac{\varepsilon_{i l}}{K^{\rm eff}_l{\rm min}(z_c,z_l)} \, ,
\label{eq:eta}
\end{align} 
where $z_c=M_N/T_c$, $T_c\sim 149$ GeV being the critical temperature below which the sphaleron transition processes freeze-out~\cite{Cline:1993bd, DOnofrio:2012jk}, $z_l\simeq 1.25 \log(25 K_l^{\rm eff})$~\cite{Deppisch:2010fr} and 
\begin{align}
K^{\rm eff}_l \ = \ \kappa_l \sum_i K_i B_{i l} \, .
\label{eq:Keff}
\end{align} 
Here the $K$-factors are defined by $K_i=\Gamma_{i}/H_N$, where $H_N=1.66 \sqrt{g_*} M_N^2/M_{\rm Pl}$ is the Hubble rate at temperature $T=M_N$, $M_{\rm Pl}=1.22\times 10^{19}$ GeV is the Planck mass and $g_*\simeq 106.75$ are the relativistic degrees of freedom at that temperature. In Eq.~\eqref{eq:Keff}, $B_{i l}$'s are the branching ratios of the $N_i$ decay to leptons of the $l$th flavor: $B_{i l}=|Y_{\nu_{i l}}|^2/(Y_\nu Y^\dag_\nu)_{ii}$. Finally, the $\kappa$-factor in Eq.~\eqref{eq:Keff} includes the effect of the real intermediate state subtracted collision terms:
\begin{align}
\kappa_l \ = \ 2\sum_{i,j \: (j\neq i)} 
\frac{{\rm Re}\left[Y_{\nu_{i l}}Y^*_{\nu_{j l}}(YY^\dag)_{ij}\right]+{\rm Im}\left[(Y_{\nu_{i l}}Y^*_{\nu_{j l}})^2\right]}{{\rm Re}\left[(Y^\dag Y)_{ll}\{(YY^\dag)_{ii}+(YY^\dag)_{jj}\} \right]}\left(1-2i\frac{M_{i}-M_{j}}{\Gamma_{i}+\Gamma_{j}}\right)^{-1} \, .
\end{align}

\begin{figure}[t]
\begin{center}
\includegraphics[width=7.7cm, angle=0]{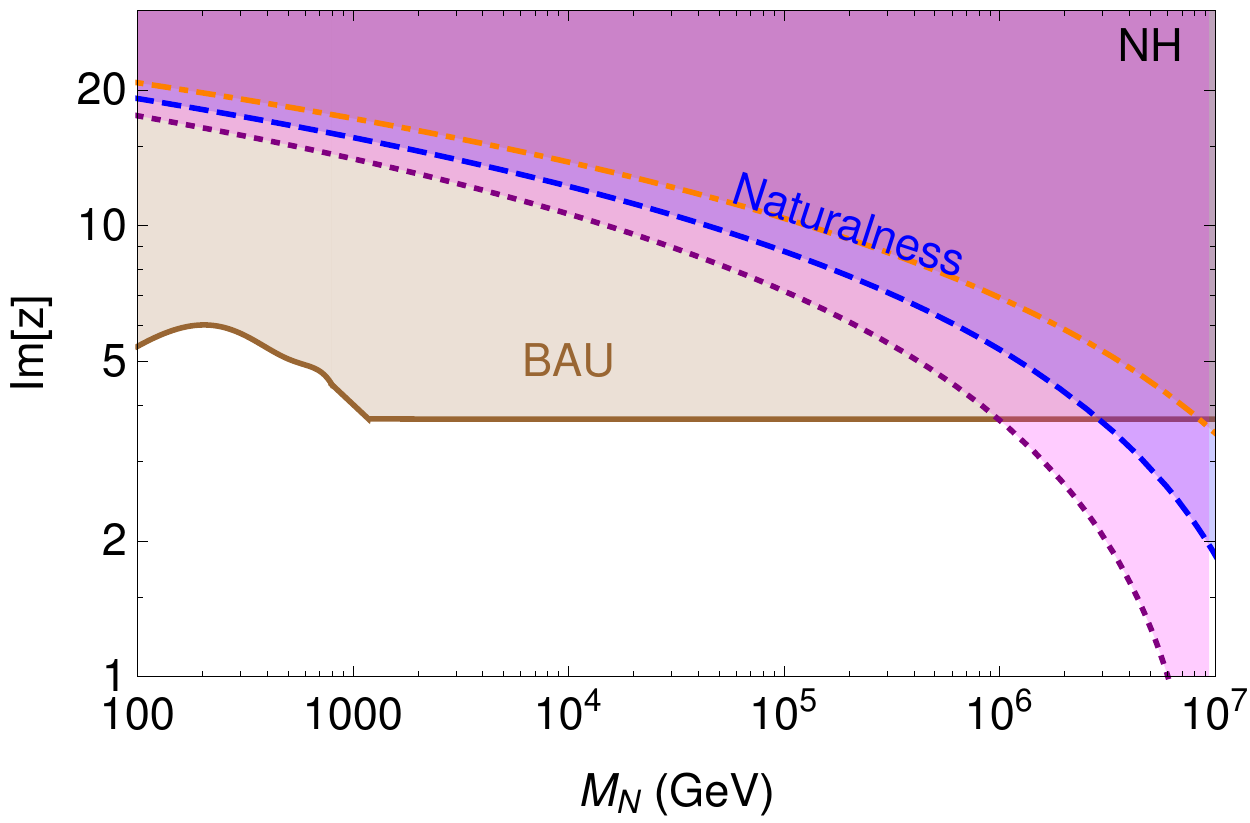} 
\caption{Leptogenesis constraints on the minimal seesaw parameter space. The region above the brown shaded region cannot reproduce the correct BAU (either in magnitude or in sign). For comparison, the naturalness constraints [cf.\ Figure~\ref{fig:Cs_mediation1}] are also shown for $\delta \mu^2 < (p~{\rm TeV})^2$, where $p=5$ (orange), 1 (blue) and 0.2 (magenta), the shaded regions being disfavored.}
\label{fig:lepto1}
\end{center}
\end{figure}

Using Eq.~\eqref{eq:eta} and the CI parametrization \eqref{ynu} for the Yukawa couplings, we calculate the BAU as a function of the average heavy-neutrino mass and the CI parameter Im[$z$]. Here we have assumed Re[$z$]=0 for simplicity, since the naturalness discussion is unaffected by this choice. The result is shown in Figure~\ref{fig:lepto1} for NH. The brown shaded region cannot reproduce the BAU within $3\sigma$ of the measured  value: $\eta_B^{\rm obs}= (6.04\pm 0.08)\times 10^{-10}$~\cite{Ade:2015xua}, either in magnitude or in sign, and is therefore disfavored. On the other hand, in the white region below it, there always exists a suitable combination of the hitherto unknown $CP$ phases $\delta$ and $\alpha$ (see Figure~\ref{fig:lepto2} below) which can reproduce the observed BAU. Note that the BAU constraint is almost independent of Im[$z$]  for $M_N\gtrsim 1~{\rm TeV}$. For lower masses closer to the electroweak scale, the observed asymmetry requires a larger value of Im[$z$]. As we go below the critical temperature $T_c$  for sphaleron transitions, the conversion efficiency for the lepton-to-baryon asymmetry drops exponentially.

Comparing this result with the other constraints shown in Figure~\ref{fig:Cs_mediation1}, we find that the leptogenesis constraints are more stringent up to $M_N\lesssim 2\times 10^6$ GeV, beyond which the naturalness constraint $\delta \mu^2<(1~{\rm TeV})^2$ (which excludes the blue shaded region) takes over.  For smaller (larger) values of $(\delta \mu^2)_{\rm max}$, the naturalness constraint will be stronger (weaker), as illustrated in Figure~\ref{fig:lepto1} for $\delta \mu^2<(p~{\rm TeV})^2$ with $p=0.2$ (magenta) and 5 (orange), respectively. %For simplicity, we do not consider the scenario with $M_N$ much smaller than the electroweak scale, where either heavy neutrino oscillations~\cite{Drewes:2016jae} or Higgs decay~\cite{Hambye:2016sby} could source the required $CP$ asymmetry. 

To illustrate the point that in the unshaded region of Figure~\ref{fig:lepto1}, there always exists a suitable combination of the $CP$ phases to reproduce the correct BAU, we have plotted the contours of correct BAU as a function of these phases in Figure~\ref{fig:lepto2}. Here we have fixed $M_N=1$ TeV, Im[$z$]=0.3 and the other PMNS parameters at their NH best-fit values for illustration. Similar plots can be produced for any other allowed values of $M_N$ and Im[$z$]. Note that a future measurement of the Dirac $CP$ phase $\delta$ with sufficient precision would imply that for a given set of values for the minimal seesaw parameters $M_N$ and Im[$z$], there exist only a finite number of choices for the Majorana phase $\alpha$ that could explain the observed BAU. One can derive similar conclusions for the IH case.   
\begin{figure}[t]
\begin{center}
\includegraphics[width=7.7cm, angle=0]{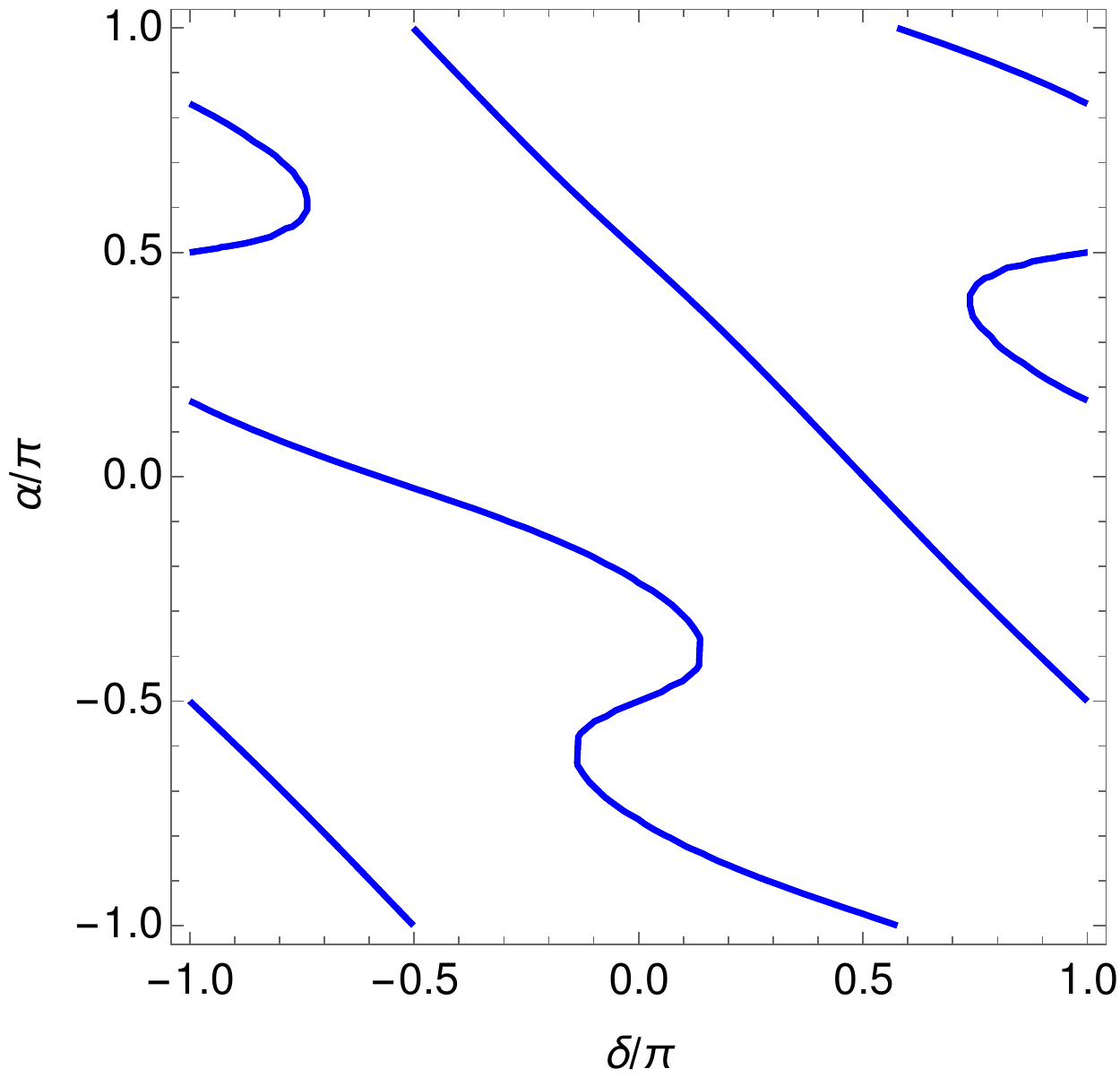}
\caption{Contours of the correct BAU as a function of the low energy $CP$ phases. Here we have fixed $M_N=1$ TeV and Im[$z$]=0.3. }
\label{fig:lepto2}
\end{center}
\end{figure}

%\newpage
\section{\label{sec:concl}Conclusion}

We have studied the implications of naturalness on the parameter space of a minimal 
type-I seesaw model. We have considered two degenerate heavy right-handed 
neutrinos and evaluated their correction to the Higgs mass 
parameter, obtaining thereby naturalness constraints on the heavy neutrino mass and Yukawa couplings. 
%We find that the right-handed neutrino masses could lie in the range 
%$10^{4} - 10^{7}$ GeV.  
We have compared these bounds with constraints stemming from 
the stability of the electroweak vacuum. 
%For larger allowed values of Yukawa couplings the 
%vacuum becomes unstable at a lower energy as compared to the Standard Model. 
%However it is possible to find parameter regions where the metastability condition is maintained. 
%Demanding this we obtain constraints on the 
%Yukawa coupling and heavy right-handed neutrino masses. 
Lepton flavor violation constraints from the decay 
$\mu \rightarrow e \gamma$ are also important for heavy neutrino masses below TeV range, whereas naturalness provides stronger constraints for masses above $10^4$ GeV or so. Metastability provides important limits in the intermediate regime. 

In the allowed parameter space we have furthermore studied the effect 
on the RG evolution of neutrino mass and mixing parameters. Although
the running shows the hallmark of threshold effects with some of the mixing 
angles showing bi-directional running from low to high scale depending on the 
parameters, in general due to the hierarchical nature of the light mass spectrum
the running is not very significant. 

We have also discussed the possibility of successful leptogenesis 
in this model. This is achieved by introducing a small mass splitting between the two heavy neutrinos comparable to their decay width. In this scenario, we find that leptogenesis provide the most stringent constraints for heavy neutrino masses below $10^6$ GeV or so, while naturalness constraints are stronger for higher masses. 

The model considered here represents the most economic seesaw scenario in terms of
particle content that can be consistent with observed neutrino masses 
in oscillation experiments, naturalness, metastability of the 
electroweak vacuum, lepton flavor violation and leptogenesis.

\section*{Acknowledgments}
We thank Serguey Petcov and Branimir Radovčić for helpful comments and discussions. B.D. was supported in part by the DFG grant RO 2516/5-1. S.K. was supported in part by UGC Dr.~D.S. Kothari Postdoctoral Fellowship grant No.~F.4-2/2006(BSR)/PH/14-15/0117. W.R. was supported by the DFG in the Heisenberg Programme with grant RO 2516/6-1.

%\newpage
%%%%%%%%%%%%%%%%%%%%%%%%%%%%%%%%%%%%%%%%%%%%%%%%%%%%%%%%%%%%%%%%%%%%%%%%%%%%%%%%%%%%%%%
%\bibliographystyle{plain}
% \bibliographystyle{unsrt}
%\bibliographystyle{apsrev4-1}
%\bibliographystyle{apsrev}
%\bibliographystyle{decsci}
\bibliography{natural_111116}
%\newpage

\end{document}